\documentclass[journal]{IEEEtran}

\usepackage{cite}
\usepackage{amsmath,amssymb,amsfonts}
\usepackage{graphicx}
\usepackage{textcomp}
\usepackage{xcolor}
\usepackage{booktabs}
\usepackage{multirow}
\usepackage{diagbox}
\usepackage{hyperref}
\usepackage{array}
\usepackage{url}
\usepackage{bm}
\usepackage{balance}
\usepackage{algorithm}
\usepackage{algorithmic}
\usepackage[caption=false,font=footnotesize]{subfig}
\usepackage{float}

\hypersetup{
 colorlinks=true,
 linkcolor=blue,
 citecolor=blue,
 urlcolor=blue,
}

\begin{document}

\title{QuaMoE-DRF: Proactive Beam and Rate Adaptation via Multimodal Dynamic Radio Map Forecasting in ISAC Networks}

\author{Zhihan Zeng, \IEEEmembership{Graduate Student Member, IEEE}, 
Kaihe Wang, \IEEEmembership{Graduate Student Member, IEEE},\\
Zhongpei Zhang, Chongwen Huang, \IEEEmembership{Senior Member, IEEE}
\thanks{Zhihan Zeng and Zhongpei Zhang are with the National Key Laboratory of Wireless Communications, University of Electronic Science and Technology of China, Chengdu 611731, China (e-mail: 202511220608@std.uestc.edu.cn; zhangzp@uestc.edu.cn). Kaihe Wang is with the University of Electronic Science and Technology of China, Chengdu 611731, China (e-mail: khewang@yeah.net). Chongwen Huang is with the College of Information Science and Electronic
Engineering, Zhejiang University, Hangzhou 310027, China; with Zhejiang
Provincial Key Laboratory of Multi-Modal Communication Networks and
Intelligent Information Processing; and also with the National Key Laboratory
of Millimeter-Wave and Terahertz Remote Sensing, Hangzhou 310027, China (emails:  chongwenhuang@zju.edu.cn). The corresponding author is Zhongpei Zhang.}}

\maketitle

\begin{abstract}
Static radio maps provide location-dependent propagation priors, but they cannot capture short-term blockage caused by moving objects. Direct sensing-assisted beam prediction is also limited because a beam index discards SINR margins, MCS thresholds, BS alternatives, and communication-equivalent neighboring beams. This paper proposes QuaMoE-DRF, a quality-aware multimodal dynamic radio map forecasting framework for proactive beam and rate adaptation in ISAC networks. Its core representation is a future beam-SINR field. We show that the full multi-BS beam-SINR field is sufficient for finite-codebook threshold-rate BS, beam, MCS, goodput, and outage decisions. For tractability, the implemented model learns a compact reference-BS local field, complemented by BS-level supervision, joint BS--beam supervision, and latent network context; we also clarify that this compact projection alone is not sufficient for BS association. QuaMoE-DRF fuses static geometry, event-like motion observations, structured sensing states, and wireless history through a quality-aware mixture-of-experts module motivated by inverse-variance fusion under heteroscedastic modality errors. It jointly predicts communication-oriented map channels and proactive BS, beam, and MCS decisions. On a dynamic multi-BS and multi-UE urban benchmark, QuaMoE-DRF achieves 402.5 Mbps effective rate, 0.0417 outage probability, and 0.1836 map RMSE, improving the effective rate by 5.67\% and reducing outage by 8.35\% over the strongest completed effective-rate baseline. The current validation uses labels from a compact blockage/path-loss simulator, with ray tracing used only for calibration and sanity checking.
\end{abstract}

\begin{IEEEkeywords}
Integrated sensing and communication, dynamic radio map, channel knowledge map, multimodal fusion, mixture of experts, dynamic blockage, beam prediction, rate adaptation.
\end{IEEEkeywords}

\section{Introduction}
\label{sec:introduction}

\IEEEPARstart{T}{he} evolution toward sixth-generation (6G) wireless networks is shifting mobile systems from connection-centric infrastructures to sensing-capable and environment-aware platforms. In the IMT-2030 vision, future networks are expected to provide not only high data rate, low latency, ubiquitous connectivity, and high reliability, but also new capabilities enabled by integrated sensing and communication (ISAC), native artificial intelligence, and environment-aware operation \cite{itur2023imt2030,liu2022towardisac,zeng2026jsrgfnet,zeng2024tutorialckm}. This shift is especially important in dense urban and vehicular high-frequency networks. Millimeter-wave propagation, large antenna arrays, narrow beams, and frequent link-state changes make conventional reactive control increasingly costly, because beam training, pilot measurements, and feedback must be repeated whenever the propagation condition changes \cite{rangan2014mmwavecellular,rappaport2017overviewmmwave,larsson2014massivemimo,zeng2026skanet,lu2014overviewmassivemimo,heath2016mmwaveoverview,giordani2019beammanagement,zeng2026phygmoe,barati2016initialaccess}. A central communication problem in such systems is therefore how to translate heterogeneous environmental observations into reliable transmission decisions before the link quality actually degrades.

Dynamic blockage is a major reason why this problem is difficult. In urban road scenes, buses, trucks, and other moving objects may temporarily block dominant line-of-sight components or strong reflected paths, producing abrupt power loss, beam misalignment, handover failure, and MCS mismatch \cite{jain2019mobileblockers,zeng2026gackan,perfecto2017v2vbeam,yan2019handoverml,khosravi2021handover}. Pilot-based channel acquisition and channel-state feedback observe the link after the channel has changed, while exhaustive or sequential beam sweeping introduces additional latency and resource overhead. Beam tracking and learning-based beam alignment reduce this cost by exploiting temporal correlation, but high mobility and short blockage duration require a predictor of future beam and link quality rather than a tracker of the current state only \cite{giordani2019beammanagement,zeng2026urbanrtrm,liu2020radarassisted,tan2024v2xsurvey}. Proactive beam and rate adaptation should therefore be built on a representation that is both predictive and communication meaningful.

Radio maps and channel knowledge maps (CKMs) provide one such representation for environment-aware communications. They store location-specific channel knowledge and have been used for coverage estimation, beam management, pilot reduction, and environment-aware hybrid beamforming \cite{debroy2015spectrummap,levie2021radiounet,zeng2021towardckm,zeng2024tutorialckm,wu2024environmentawareckm}. Recent task-oriented radio-map and CKM methods further demonstrate that learned propagation maps can assist beamforming, reduce pilot overhead, reconstruct missing map regions, and support aerial or three-dimensional networks \cite{dai2024ckmprototype,yang2025reducedpilots,zhang2024radiomapinpainting,chen2024diffractionradiomap,zhan2024aerialvideoeckm}. Generative methods have also been introduced to improve radio-map fidelity under sparse or degraded observations \cite{zhang2023rmegan,luo2025ddpmradiomap}. Most of these maps, however, are static or quasi-static. They capture building-induced shadowing and long-term spatial regularities, but they do not indicate whether a moving vehicle will cross a serving BS--UE path in the next few slots. Thus, a static radio map may be accurate for the background environment and still be insufficient for blockage-aware link adaptation.

Sensing-assisted communication addresses the dynamic part of the problem by observing mobile objects and estimating their motion. Radar-assisted predictive beamforming, LiDAR-aided beam selection, vision-aided beam alignment, camera-based beam prediction, and semantic sensing have shown that environmental observations can reduce beam training overhead and improve robustness in high-mobility scenarios \cite{liu2020roadisac,zhang2021isacsurvey,liu2020radarassisted,klautau2019lidarbeamselection,jiang2023lidarfuturebeam,xu2022visionbeamalignment}. Large-scale multimodal sensing and communication datasets further make it possible to combine wireless measurements with camera, LiDAR, radar, and semantic information for future beam prediction and blockage-aware handover \cite{alkhateeb2023deepsense6g,ahn2022visionbeamselection,charan2021visionaided,charan2023camerabeam,yang2023semanticsjsac,patel2024multimodalsensing,mollah2026multimodality}. Nevertheless, sensing is not by itself a communication decision. Object positions and velocities do not directly specify which blocker is critical to a BS--UE path, which BS should serve the UE, which beam remains robust, or which MCS is supportable under the realized SINR. In addition, modality reliability changes with traffic density, sensing noise, visibility, blockage severity, and recent link quality. Geometry is stable but blind to moving blockers; event-like observations emphasize motion but can be sparse; structured sensing gives object-level kinematics but may be incomplete; wireless history is directly communication related but can lag behind the future blockage event. A practical predictor must therefore learn not only what to fuse, but also when each modality should be trusted.

Motivated by this gap, this paper studies proactive beam and rate adaptation through multimodal dynamic radio map forecasting in ISAC networks. The key design choice is the intermediate representation. A direct beam label is compact, but it is not a complete communication state: it does not tell how far the selected beam is from an MCS threshold, whether another BS has a safer margin, or whether a neighboring beam is rate-equivalent. At the theoretical level, the full multi-BS beam-SINR field is a sufficient intermediate state for the finite-codebook and threshold-MCS decision rule, because the oracle BS, beam, MCS, goodput, outage, and tied alternatives are deterministic functions of that field. The implemented target is deliberately more compact: it stores the beam-SINR slice of the reference serving BS on the future local grid and combines this slice with BS-level decision supervision, joint BS--beam supervision, and latent network context. This distinction avoids treating the compact map alone as a sufficient statistic for BS association, while still preserving the margins needed for proactive beam and rate decisions at much lower dimension.

The second design issue is reliability. Multimodal fusion is useful only when the fusion rule changes with the failure regime. Event maps may be missing, object-level sensing may be noisy or partially occluded, and wireless history may become stale when a blocker enters the link after the last pilot. A fixed concatenation assigns the same trust pattern in all these regimes. QuaMoE-DRF therefore uses quality-aware mixture-of-experts routing, where reliability cues modulate both expert selection and modality weighting. We show that, under a standard heteroscedastic error model, the optimal linear fusion weight is inverse-variance dependent and therefore state dependent. This gives a communication-theoretic reason for quality-aware routing beyond the fact that MoE is a flexible neural module.

The main contributions are summarized as follows.
\begin{itemize}
    \item We formulate proactive beam and rate adaptation as dynamic beam-SINR radio map forecasting. The full multi-BS beam-SINR field is shown to be sufficient for finite-codebook threshold-rate BS, beam, MCS, goodput, and outage decisions, while the compact reference-BS projection is proved insufficient for BS association alone.

    \item We derive margin-based reliability relations that link SINR-map prediction error, modality uncertainty, MCS margins, and prediction horizon to goodput and outage. The analysis explains why errors near MCS thresholds, beam boundaries, and dynamic blockage transitions are most harmful.

    \item We propose QuaMoE-DRF, a quality-aware multimodal forecasting framework that fuses static geometry, event-like motion, structured sensing states, and wireless history. Its event-sensing calibration and sparse expert routing are motivated by heteroscedastic inverse-uncertainty fusion.

    \item We evaluate QuaMoE-DRF on a dynamic multi-BS and multi-UE urban benchmark. The results show improvements in effective rate, outage probability, beam robustness, map accuracy, blockage estimation, and robustness under modality degradation, with label provenance and simulator limitations explicitly reported.
\end{itemize}

The remainder of this paper is organized as follows. Section~\ref{sec:system_model} introduces the system model and formulates the proactive dynamic radio map forecasting problem. Section~\ref{sec:method} presents the proposed QuaMoE-DRF framework and the main theoretical results. Section~\ref{sec:setup} describes the benchmark, compared baselines, and evaluation metrics. Section~\ref{sec:results} reports and analyzes the simulation results. Section~\ref{sec:conclusion} concludes the paper.

\section{System Model and Problem Formulation}
\label{sec:system_model}

\subsection{Dynamic Urban ISAC Network}

\begin{figure}[htbp]
    \centering
    \includegraphics[width=\columnwidth]{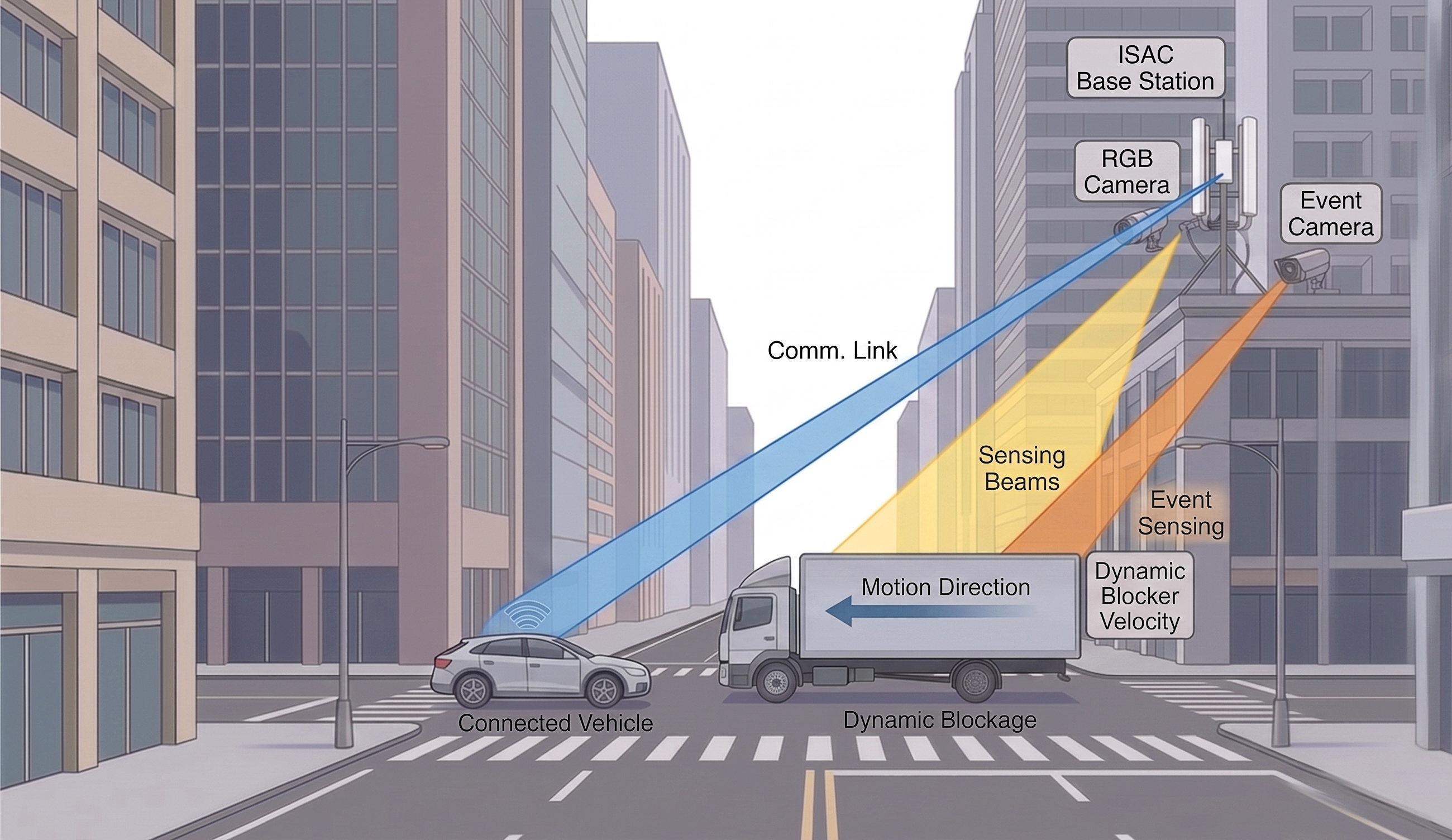}
    \caption{Dynamic urban ISAC scenario for blockage-aware proactive beam and rate adaptation.}
    \label{fig:architecture}
\end{figure}

We consider an ISAC-enabled downlink network deployed in a dynamic urban road scene, as shown in Fig.~\ref{fig:architecture}. The horizontal service region is denoted by $\mathcal{D}\subset\mathbb{R}^2$ and is represented by a rasterized scene map. The candidate BS set is $\mathcal{B}=\{1,\ldots,M\}$, where BS $m$ is located at $\boldsymbol{b}_m\in\mathbb{R}^2$. Each BS is equipped with a predefined beam codebook $\mathcal{F}=\{\boldsymbol{f}_1,\ldots,\boldsymbol{f}_K\}$. The communication decision is made at slot $t$ for a future slot $t+\Delta$. The formulation is written for one scheduled UE for notational clarity; the same procedure is applied to every scheduled UE in the benchmark. The UE position at slot $t$ is denoted by $\boldsymbol{u}(t)\in\mathbb{R}^2$, and the scene contains moving blockers such as vehicles. For blocker $i$, the observable state is represented by its center position, velocity, and spatial extent. The future local map is defined on a grid $\Omega_{t+\Delta}$ centered at the kinematic prediction of the UE position obtained from the current UE state rather than from future communication labels. This formulation allows the model to predict a local future dynamic radio map without using future BS, beam, MCS, rate, or blockage labels as inputs.
\begin{align}
    \boldsymbol{o}_i(t)
    &=
    \left[
    x_i(t),y_i(t),
    v_{x,i}(t),v_{y,i}(t),
    \ell_i(t),w_i(t)
    \right]^T, \\
    \Delta
    &=N_pT_s, \\
    \Omega_{t+\Delta}
    &=
    \left\{\boldsymbol{x}_{r,c}\right\}_{r=1,c=1}^{H_l,W_l}.
    \label{eq:scene_grid}
\end{align}

\subsection{Blockage-Aware Link and Beam Model}

For a candidate BS $m$ and a point $\boldsymbol{x}\in\mathcal{D}$, the link geometry is determined by distance and azimuth. The large-scale received power is modeled by a distance-dependent path loss term and additional attenuation caused by static structures, moving blockers, and invalid building interior regions. The model is deliberately compact: it keeps the geometric path loss and the static and dynamic attenuation terms that dominate BS association, beam selection, and rate adaptation in the considered benchmark. It should therefore be interpreted as a controlled communication benchmark rather than as a replacement for full three-dimensional ray tracing.
\begin{align}
    d_m(\boldsymbol{x})
    &=
    \left\|\boldsymbol{x}-\boldsymbol{b}_m\right\|_2, \\
    \phi_m(\boldsymbol{x})
    &=
    \operatorname{atan2}
    \left(
    [\boldsymbol{x}-\boldsymbol{b}_m]_2,
    [\boldsymbol{x}-\boldsymbol{b}_m]_1
    \right), \\
    L_m(\boldsymbol{x})
    &=
    L_0
    +10\alpha\log_{10}
    \left(
    \frac{\max\{d_m(\boldsymbol{x}),d_0\}}{d_0}
    \right), \\
    \begin{split}
    P^{\rm rx}_{m}(\boldsymbol{x},t)
    &=
    P^{\rm tx}
    -L_m(\boldsymbol{x})
    -\lambda_{\rm sta}a_m^{\rm sta}(\boldsymbol{x})
    \\
    &\quad
    -\lambda_{\rm dyn}a_m^{\rm dyn}(\boldsymbol{x},t)
    -\lambda_{\rm in}\mathbf{1}\{\boldsymbol{x}\in\mathcal{I}\}.
    \end{split}
    \label{eq:large_scale_power}
\end{align}
Here $L_0$, $d_0$, and $\alpha$ are the reference path loss parameters, $a_m^{\rm sta}$ and $a_m^{\rm dyn}$ are the static and dynamic occlusion scores, $\lambda_{\rm sta}$ and $\lambda_{\rm dyn}$ are the corresponding attenuation factors, and $\mathcal{I}$ denotes the building interior region. The $k$th beam has central angle $\bar\phi_k$ and beam width $\Phi_{\rm bw}$. Its directional gain is expressed in the dB domain as $g_k(\phi)$. The beam-wise received power and SINR are also written in the dB domain, which is consistent with the MCS thresholds used for link adaptation.
\begin{align}
    g_k(\phi)
    &=
    \operatorname{clip}
    \left[
    -12
    \left(
    \frac{\angle(\phi-\bar\phi_k)}{\Phi_{\rm bw}}
    \right)^2,
    -18,0
    \right], \\
    P^{\rm rx}_{m,k}(\boldsymbol{x},t)
    &=
    P^{\rm rx}_{m}(\boldsymbol{x},t)
    +g_k(\phi_m(\boldsymbol{x})), \\
    \gamma^{\rm dB}_{m,k}(\boldsymbol{x},t)
    &=
    P^{\rm rx}_{m,k}(\boldsymbol{x},t)
    -10\log_{10}
    \left(
    10^{P_N/10}+10^{P_I/10}
    \right).
    \label{eq:beam_sinr}
\end{align}

\subsection{MCS-Constrained Communication Objective}

Let $\mathcal{C}=\{0,\ldots,C-1\}$ denote the MCS set. MCS $c$ has dB threshold $\Gamma_c$ and spectral efficiency $\eta_c$. We additionally use $c=-1$ as an outage/no-service index with $\eta_{-1}=0$ and $\Gamma_{-1}=-\infty$. For a selected BS, beam, and MCS, the realized rate follows a threshold-based link adaptation rule. The oracle decision is generated by first maximizing the MCS-constrained rate and then using SINR to resolve ties. If no positive MCS threshold is satisfied, the link is assigned to the outage state and the realized rate is set to zero. This rule is important because several neighboring beams may support the same MCS and therefore produce communication-equivalent decisions.
\begin{equation}
\begin{aligned}
    c^{\star}_{m,k}(\boldsymbol{x},t)
    =
    \max\Big(
    \{-1\}
    \cup
    \{c\in\mathcal{C}:\gamma^{\rm dB}_{m,k}(\boldsymbol{x},t)\ge\Gamma_c\}
    \Big).
\end{aligned}
\label{eq:mcs_index}
\end{equation}
The inclusion of $\{-1\}$ makes the maximum well-defined even when the feasible MCS set is empty.
\begin{equation}
    R_{m,k,c}(\boldsymbol{x},t)
    =
    B_0\eta_c
    \mathbf{1}
    \{c=-1\ \text{or}\ \gamma^{\rm dB}_{m,k}(\boldsymbol{x},t)\ge\Gamma_c\},
\label{eq:threshold_rate}
\end{equation}
where $R_{m,k,-1}=0$ because $\eta_{-1}=0$.
\begin{equation}
    R^{\star}_{m,k}(\boldsymbol{x},t)
    =
    B_0\eta_{c^{\star}_{m,k}(\boldsymbol{x},t)}.
\label{eq:oracle_link_rate}
\end{equation}
\begin{equation}
\begin{aligned}
    (m^{\star},k^{\star})
    &=
    \underset{m\in\mathcal{B},\,k\in\{1,\ldots,K\}}{\arg\max}\;
    \Big(
    R^{\star}_{m,k}(\boldsymbol{x},t),
    \gamma^{\rm dB}_{m,k}(\boldsymbol{x},t)
    \Big).
\end{aligned}
\label{eq:oracle_decision}
\end{equation}
The maximization in \eqref{eq:oracle_decision} is lexicographic. The rate is optimized first, and the SINR is used only when two candidate decisions provide the same MCS-constrained rate. Therefore, the later evaluation reports both strict beam accuracy and communication-oriented beam metrics such as Beam@3 and tied beam accuracy.

For a predicted decision $(\hat m,\hat k,\hat c)$, the outage indicator used in evaluation is
\begin{equation}
    O(\hat m,\hat k,\hat c)
    =
    \mathbf{1}\{\hat c=-1\}
    +
    \mathbf{1}\{\hat c\in\mathcal{C},\gamma^{\rm dB}_{\hat m,\hat k}<\Gamma_{\hat c}\},
    \label{eq:outage_indicator}
\end{equation}
where the two events are mutually exclusive. Thus, a deliberate no-service decision and an unsupported positive-MCS decision are both counted as outage in the reported outage probability.

\subsection{Multimodal Observation Model}

At slot $t$, the predictor observes static geometry, event-like motion information, structured sensing states, wireless history, and modality quality cues. The geometry tensor describes road layout, building occupancy, obstacle support, and BS locations. The event tensor describes motion saliency and dynamic occupancy changes. The sensing vector stacks observed blocker states, UE kinematic information, relative geometry, and padded network context. The wireless history tensor stores recent beam and link-quality observations. All wireless history entries are obtained from slots no later than $t$, and they do not include future BS, beam, MCS, blockage, or rate labels. The quality vector provides soft reliability cues for different modalities and is used for fusion rather than hard modality selection.
\begin{align}
    \mathcal{O}_t
    &=
    \left\{
    \mathbf{X}^{\rm geo},
    \mathbf{X}^{\rm evt}_{t},
    \boldsymbol{x}^{\rm sen}_{t},
    \mathbf{X}^{\rm wir}_{t-T_h+1:t},
    \boldsymbol{q}_{t}
    \right\}, \\
    \boldsymbol{q}_t
    &=
    \left[
    q_{\rm geo},
    q_{\rm evt},
    q_{\rm sen},
    q_{\rm wir}
    \right]^T.
    \label{eq:observation_model}
\end{align}

\begin{figure}[htbp]
    \centering
    \includegraphics[width=\columnwidth]{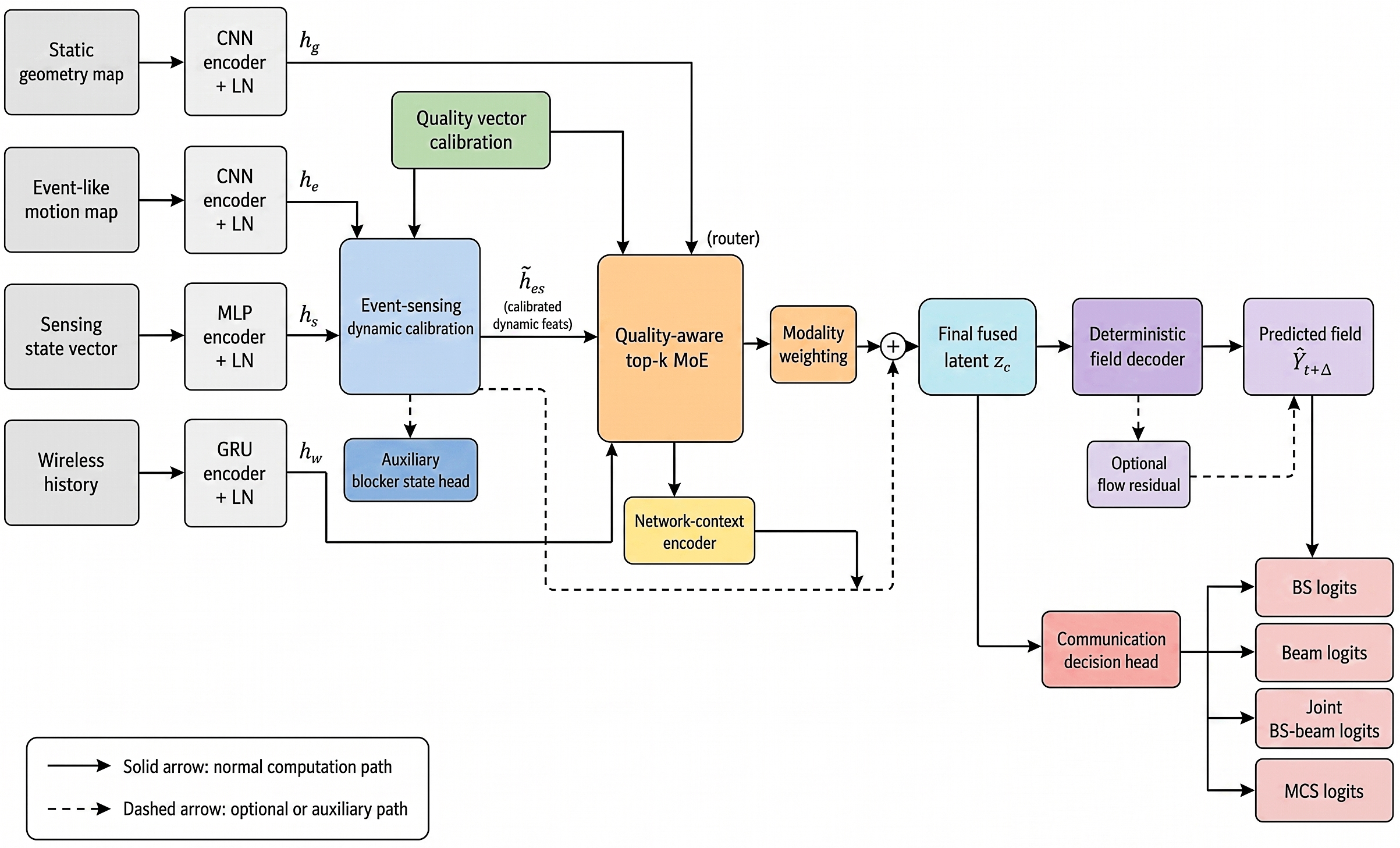}
    \caption{Architecture of the proposed QuaMoE-DRF framework with multimodal encoding, event-sensing calibration, quality-aware MoE fusion, dynamic field prediction, and communication decision heads.}
    \label{fig:quamoe_architecture}
\end{figure}

\subsection{Future Dynamic Radio Map Target}

The prediction target is a future local dynamic radio map rather than a purely geometric or visual reconstruction. For each grid point in $\Omega_{t+\Delta}$, the simulator first evaluates the full candidate BS--beam SINR field using the blockage-aware link model:
\begin{align}
    \mathbf{\Gamma}_{t+\Delta}
    &=
    \left[
    \gamma^{\rm dB}_{m,k}(\boldsymbol{x}_{r,c},t+\Delta)
    \right]_{m,k,r,c}, \\
    \mathbf{\Gamma}_{t+\Delta}
    &\in
    \mathbb{R}^{M\times K\times H_l\times W_l}.
    \label{eq:full_bssinr_field}
\end{align}
This full multi-BS field is the communication-sufficient object formalized in Section~\ref{sec:method} and proved in Appendix~\ref{app:proofs}. The implemented target, however, does not store all $M K$ beam-SINR channels. To reduce dimension, it stores a compact reference-BS projection of this field, together with one normalized received-power channel, $K$ beam-wise link-quality channels, one blockage-likelihood channel, and one communication-ambiguity channel. The reference BS is selected by the same label-generation rule used for the future local decision, while BS association is learned through separate BS and joint BS--beam decision labels.
\begin{align}
    \mathbf{Y}_{t+\Delta}
    &=
    \Pi_{\rm ref}(\mathbf{\Gamma}_{t+\Delta},m^{\star}), \\
    \mathbf{Y}_{t+\Delta}
    &=
    \left[
    \mathbf{Y}^{\beta}_{t+\Delta},
    \mathbf{Y}^{1}_{t+\Delta},
    \ldots,
    \mathbf{Y}^{K}_{t+\Delta},
    \mathbf{Y}^{\rm blk}_{t+\Delta},
    \mathbf{Y}^{\rm unc}_{t+\Delta}
    \right], \\
    \mathbf{Y}_{t+\Delta}
    &\in
    \mathbb{R}^{(K+3)\times H_l\times W_l}.
    \label{eq:future_map}
\end{align}
The compact map should therefore be read as a low-dimensional surrogate of the full sufficient field, not as a stand-alone sufficient statistic for BS association. BS-level supervision and latent network context carry the missing inter-BS information, whereas the compact map preserves the serving-link beam margins that determine beam robustness and MCS support.

For completeness, the normalization used in all map channels is defined as
\begin{equation}
    \operatorname{norm}_{[a,b]}(z)
    \triangleq
    \operatorname{clip}\left(\frac{z-a}{b-a},0,1\right),
    \quad a<b.
    \label{eq:norm_definition}
\end{equation}
With received-power normalization constants $P_{\min}$ and $P_{\max}$ and SINR normalization constants $\Gamma_{\min}$ and $\Gamma_{\max}$, the first $K+1$ channels are
\begin{align}
    Y^{\beta}_{r,c}
    &=
    \operatorname{norm}_{[P_{\min},P_{\max}]}
    \left(P^{\rm rx}_{m^{\star}}(\boldsymbol{x}_{r,c},t+\Delta)\right), \\
    Y^{k}_{r,c}
    &=
    \operatorname{norm}_{[\Gamma_{\min},\Gamma_{\max}]}
    \left(\gamma^{\rm dB}_{m^{\star},k}(\boldsymbol{x}_{r,c},t+\Delta)\right),
    \quad k=1,\ldots,K.
    \label{eq:normalized_map_channels}
\end{align}
In the experiments, $\Gamma_{\min}$ and $\Gamma_{\max}$ are chosen to cover the MCS threshold range with margin. The blockage channel is written as a blockage likelihood because it is generated from occlusion scores rather than from a calibrated posterior probability. The ambiguity channel is also used as a communication ambiguity proxy rather than a calibrated Bayesian variance. It combines physical blockage risk and beam ambiguity, where beam ambiguity is measured by the SINR margin between the best and second-best reference-BS beams at the grid point:
\begin{align}
    Y^{\rm blk}_{r,c}
    &=
    \operatorname{clip}
    \left(
    \max
    \left\{
    a_{m^{\star}}^{\rm sta}(\boldsymbol{x}_{r,c}),
    a_{m^{\star}}^{\rm dyn}(\boldsymbol{x}_{r,c},t+\Delta)
    \right\},
    0,1
    \right), \\
    \delta\gamma_{r,c}
    &=
    \max_{k}\gamma^{\rm dB}_{m^{\star},k}(\boldsymbol{x}_{r,c},t+\Delta)
    -
    \max_{k\ne k_1}\gamma^{\rm dB}_{m^{\star},k}(\boldsymbol{x}_{r,c},t+\Delta), \\
    k_1
    &=
    \operatorname*{arg\,max}_k\gamma^{\rm dB}_{m^{\star},k}(\boldsymbol{x}_{r,c},t+\Delta), \\
    Y^{\rm unc}_{r,c}
    &=
    \operatorname{clip}
    \left[
    \frac{1}{2}
    \left(
    1-\operatorname{norm}_{[0,12]}(\delta\gamma_{r,c})
    \right)
    +\frac{1}{2}Y^{\rm blk}_{r,c},
    0,1
    \right].
    \label{eq:blockage_uncertainty}
\end{align}
For the benchmark setting $K=8$, the second maximum is always defined. If a single-beam special case is considered, we set $\delta\gamma_{r,c}=12$ dB so that the beam-ambiguity component is zero after normalization.

\subsection{Learning Problem}

The proposed predictor maps current observations to a future communication map and then produces proactive BS, beam, joint BS--beam, and MCS decisions for the future slot. The communication goal is to maximize the expected realized rate and reduce the outage probability at $t+\Delta$. Since the realized rate contains threshold operations and discrete decisions, training uses a supervised multi-task surrogate that combines map prediction, blockage likelihood estimation, uncertainty regression, and communication decision supervision. The final evaluation always follows the realized rate rule in \eqref{eq:threshold_rate} and the outage rule in \eqref{eq:outage_indicator}; hence the learning objective provides differentiable supervision while the reported performance remains communication-oriented.
\begin{align}
    \hat{\mathbf{Y}}_{t+\Delta}
    &=
    f_{\theta}(\mathcal{O}_t), \\
    (\hat m,\hat k,\hat j,\hat c)
    &=
    h_{\theta}
    \left(
    \hat{\mathbf{Y}}_{t+\Delta},
    \mathcal{O}_t
    \right), \\
    \max_{\theta}\quad
    &\mathbb{E}
    \left[
    R_{\hat m,\hat k,\hat c}
    \left(
    \boldsymbol{u}(t+\Delta),t+\Delta
    \right)
    \right], \\
    \min_{\theta}\quad
    &\Pr
    \left[
    O(\hat m,\hat k,\hat c)=1
    \right].
    \label{eq:learning_problem}
\end{align}

\section{Proposed Method}
\label{sec:method}

\subsection{Overview}

QuaMoE-DRF converts the observation set in \eqref{eq:observation_model} into a future local dynamic radio map and then into proactive BS, beam, and MCS decisions. As illustrated in Fig.~\ref{fig:quamoe_architecture}, each modality is first processed by a dedicated encoder, the encoded features are fused by a quality-aware mixture-of-experts module, and the fused latent representation is decoded into the communication field in \eqref{eq:future_map}. Event-sensing calibration, a network-context residual, and a non-leaking kinematic prior are used to strengthen the dynamic representation, while all future oracle BS, beam, MCS, blockage, and rate labels remain excluded from the inputs.

\subsection{Communication Representation: Full Sufficiency and Compact Approximation}
\label{subsec:theory_interface}

The radio-map decoder maps the fused latent representation to the future dynamic radio map tensor defined in Section~\ref{sec:system_model}. The predicted dynamic radio map is written as
\begin{align}
    \hat{\mathbf{Y}}_{t+\Delta}
    &=
    D_{\theta}(\boldsymbol{z}_c), \\
    \hat{\mathbf{Y}}_{t+\Delta}
    &=
    \left[
    \hat{\mathbf{Y}}^{\beta}_{t+\Delta},
    \hat{\mathbf{Y}}^{1}_{t+\Delta},
    \ldots,
    \hat{\mathbf{Y}}^{K}_{t+\Delta},
    \hat{\mathbf{Y}}^{\rm blk}_{t+\Delta},
    \hat{\mathbf{Y}}^{\rm unc}_{t+\Delta}
    \right], \\
    \hat{\mathbf{Y}}_{t+\Delta}
    &\in
    \mathbb{R}^{(K+3)\times H_l\times W_l}.
    \label{eq:decoder_field_quamoe}
\end{align}
The important point is that \eqref{eq:decoder_field_quamoe} is not used as a generic image reconstruction target. The full multi-BS beam-SINR field in \eqref{eq:full_bssinr_field} deterministically specifies the oracle BS, beam, MCS, goodput, and outage under the finite-codebook threshold-rate model.

\noindent\textbf{Theorem 1 (Full-field decision sufficiency).}
For a fixed future grid point $\boldsymbol{x}$, finite BS set $\mathcal{B}$, finite beam codebook $\{1,\ldots,K\}$, and threshold-MCS rule \eqref{eq:mcs_index}--\eqref{eq:oracle_decision}, the full beam-SINR vector
\begin{equation}
    \boldsymbol{\gamma}_{\boldsymbol{x}}
    =
    \{\gamma^{\rm dB}_{m,k}(\boldsymbol{x},t+\Delta):m\in\mathcal{B},k=1,\ldots,K\}
\end{equation}
is sufficient for determining the oracle BS, beam, MCS, realized rate, outage indicator, and tied communication-equivalent beam set. In particular, all these quantities are deterministic functions of $\boldsymbol{\gamma}_{\boldsymbol{x}}$.

\noindent\textbf{Proposition 1 (Non-sufficiency of compact reference projection).}
The compact reference-BS projection $\Pi_{\rm ref}(\mathbf{\Gamma}_{t+\Delta},m^{\star})$ in \eqref{eq:future_map} is not, by itself, a sufficient statistic for multi-BS association. There exist two full fields with the same compact projected serving slice but different oracle serving BSs.

The implemented map in \eqref{eq:decoder_field_quamoe} is therefore the compact reference-BS projection of a sufficient full state. It preserves beam-wise margins, blockage risk, and ambiguity around the predicted serving region, while inter-BS alternatives are learned through BS logits, joint BS--beam logits, and the latent network-context residual. This is why a beam-index label is too weak, but also why the compact map is not claimed to recover BS association by itself.

Let $\boldsymbol{z}_c$ denote the final fused latent representation. The decision head uses the latent feature, the predicted dynamic radio map, and the current sensing state to produce communication decision logits:
\begin{align}
    \{
    \boldsymbol{\ell}_{\rm BS},
    \boldsymbol{\ell}_{\rm beam},
    \boldsymbol{\ell}_{\rm joint},
    \boldsymbol{\ell}_{\rm MCS}
    \}
    =
    H_{\theta}
    \left(
    \boldsymbol{z}_c,
    \hat{\mathbf{Y}}_{t+\Delta},
    \boldsymbol{x}^{\rm sen}_{t}
    \right).
    \label{eq:decoder_decision_quamoe}
\end{align}
Here $\boldsymbol{\ell}_{\rm BS}$, $\boldsymbol{\ell}_{\rm beam}$, $\boldsymbol{\ell}_{\rm joint}$, and $\boldsymbol{\ell}_{\rm MCS}$ denote the logits for BS selection, beam selection, joint BS--beam selection, and MCS prediction. A non-leaking kinematic prior is formed from the current UE state, current velocity information, and BS geometry. Future oracle BS, beam, MCS, and rate labels are never used as inputs; they are only used as supervision and for final communication evaluation.

\noindent\textbf{Theorem 2 (Margin outage and goodput bound).}
Let $\hat\ell=(\hat m,\hat k)$ and $\hat c\in\mathcal{C}$ be selected from a predicted SINR field. Define the selected-link prediction error $\delta_{\hat\ell}=\hat\gamma_{\hat\ell}-\gamma_{\hat\ell}$ and the predicted MCS margin $\mu_{\hat\ell,\hat c}=\hat\gamma_{\hat\ell}-\Gamma_{\hat c}$. If $\delta_{\hat\ell}$ is zero-mean sub-Gaussian with variance proxy $\sigma_{f,\hat\ell}^{2}$, then
\begin{equation}
    \Pr\{\gamma_{\hat\ell}<\Gamma_{\hat c}\}
    \le
    \exp\left(-\frac{[\mu_{\hat\ell,\hat c}]_+^2}{2\sigma_{f,\hat\ell}^{2}}\right),
    \label{eq:main_margin_outage_bound}
\end{equation}
and the expected goodput satisfies
\begin{equation}
    \mathbb{E}[R_{\hat\ell,\hat c}]
    \ge
    B_0\eta_{\hat c}
    \left[1-
    \exp\left(-\frac{[\mu_{\hat\ell,\hat c}]_+^2}{2\sigma_{f,\hat\ell}^{2}}\right)
    \right].
    \label{eq:main_margin_goodput_bound}
\end{equation}
If only a finite second moment is assumed, the same structure holds with Cantelli's bound, as detailed in Appendix~\ref{app:proofs}.

Theorem~2 connects map error to communication loss. A low RMSE is most valuable when it reduces errors near MCS thresholds and blockage boundaries; conversely, a small average map error can still cause outage if it occurs at a small-margin selected link.

\noindent\textbf{Theorem 3 (Heteroscedastic inverse-uncertainty fusion).}
Suppose that modality-specific local SINR estimates satisfy $\hat\gamma_{\ell}^{(i)}=\gamma_{\ell}+\varepsilon_{i,\ell}$, $\mathbb{E}[\varepsilon_{i,\ell}|q_i]=0$, $\operatorname{var}(\varepsilon_{i,\ell}|q_i)=\sigma_{i,\ell}^2$, and $\operatorname{cov}(\varepsilon_{i,\ell},\varepsilon_{j,\ell}|q_i,q_j)=0$ for $i\ne j$. Among all unbiased linear estimators $\hat\gamma_{\ell}=\sum_i\rho_i\hat\gamma_{\ell}^{(i)}$ with $\sum_i\rho_i=1$, the minimum-variance weights are
\begin{equation}
    \rho_i^{\star}
    =
    \frac{\sigma_{i,\ell}^{-2}}
    {\sum_j\sigma_{j,\ell}^{-2}},
    \quad
    \sigma_{f,\ell}^{2,\star}
    =
    \left(\sum_j\sigma_{j,\ell}^{-2}\right)^{-1}.
    \label{eq:main_inverse_variance}
\end{equation}

Since the modality variances $\sigma_{i,\ell}^{2}$ depend on quality scores, blockage state, sensing geometry, and prediction horizon, a fixed fusion rule is generally suboptimal. The quality-aware MoE module is therefore not used merely as a generic neural block; it parameterizes a state-dependent estimator motivated by Theorem~3.

\subsection{Quality-Aware Multimodal Encoding}

QuaMoE-DRF uses four input modalities. Static geometry and event-like motion maps are encoded by CNN encoders, the structured sensing vector is encoded by an MLP encoder, and wireless history is encoded by a GRU encoder. Layer normalization is applied to the encoded features before fusion:
\begin{align}
    \boldsymbol{z}_{g}
    &=
    \operatorname{LN}
    \left(
    e_g(\mathbf{X}^{\rm geo})
    \right), \\
    \boldsymbol{z}_{e}
    &=
    \operatorname{LN}
    \left(
    e_e(\mathbf{X}^{\rm evt}_{t})
    \right), \\
    \boldsymbol{z}_{s}
    &=
    \operatorname{LN}
    \left(
    e_s(\boldsymbol{x}^{\rm sen}_{t})
    \right), \\
    \boldsymbol{z}_{w}
    &=
    \operatorname{LN}
    \left(
    e_w(\mathbf{X}^{\rm wir}_{t-T_h+1:t})
    \right).
    \label{eq:modality_encoding_quamoe}
\end{align}
The static geometry representation provides road layout, building occupancy, BS locations, and long-term spatial priors. The event representation captures motion saliency and dynamic occupancy changes. The sensing representation contains object-level states, UE kinematics, relative geometry, and padded scene context. The wireless-history representation summarizes recent link observations. Since these modalities have different failure modes, QuaMoE-DRF does not concatenate them with fixed weights. Instead, it uses quality-aware routing and modality weighting.

\subsection{Event-Sensing Dynamic Calibration}

The event-like motion map and structured sensing vector describe moving objects from complementary perspectives. The event branch provides spatial motion evidence, while the sensing branch provides object-level kinematic states. QuaMoE-DRF uses an event-sensing dynamic calibration module to update the event and sensing latent representations and to form a dynamic residual feature:
\begin{align}
    \left(
    \bar{\boldsymbol{z}}_{e},
    \bar{\boldsymbol{z}}_{s},
    \boldsymbol{d}_{es}
    \right)
    =
    C_{\theta}
    \left(
    \boldsymbol{z}_{e},
    \boldsymbol{z}_{s}
    \right),
    \label{eq:event_sensing_calibration}
\end{align}
where $\bar{\boldsymbol{z}}_{e}$ and $\bar{\boldsymbol{z}}_{s}$ are the calibrated event and sensing features, and $\boldsymbol{d}_{es}$ is the dynamic residual representation. This calibration helps align motion saliency with blocker kinematics and supports dynamic blockage-aware radio map forecasting.

\subsection{Quality-Aware Mixture-of-Experts Fusion}

For each modality, the expert router receives the modality feature and its reliability score. The router produces sparse top-$k$ expert weights, and the selected experts transform the modality feature. This process is written as
\begin{align}
    \boldsymbol{\alpha}_i
    &=
    \operatorname{TopK}
    \left(
    \operatorname{softmax}
    \left(
    r_i([\boldsymbol{z}_i,q_i])/\tau
    \right),
    k
    \right), \\
    \tilde{\boldsymbol{z}}_i
    &=
    \sum_{j=1}^{N_{\rm exp}}
    \alpha_{ij}
    E_j(\boldsymbol{z}_i),
    \quad
    i\in\{g,e,s,w\},
    \label{eq:expert_routing_quamoe}
\end{align}
where $\boldsymbol{z}_i$ denotes the modality feature, $q_i$ denotes its quality score, $r_i(\cdot)$ denotes the modality-specific router, $E_j(\cdot)$ denotes the $j$th expert, $\tau$ is the routing temperature, and $N_{\rm exp}$ is the number of experts. The operator $\operatorname{TopK}$ sets all but the largest $k$ entries to zero and renormalizes the remaining entries to sum to one.

After expert routing, modality-level weights are computed to obtain the MoE fused feature:
\begin{align}
    \boldsymbol{a}_{\rho}
    &=
    \left[
    a_g(\tilde{\boldsymbol{z}}_g,q_g),
    a_e(\tilde{\boldsymbol{z}}_e,q_e),
    a_s(\tilde{\boldsymbol{z}}_s,q_s),
    a_w(\tilde{\boldsymbol{z}}_w,q_w)
    \right]^T, \\
    \boldsymbol{\rho}
    &=
    \operatorname{softmax}(\boldsymbol{a}_{\rho}), \\
    \boldsymbol{z}_{f}
    &=
    P_f
    \left(
    \sum_{i\in\{g,e,s,w\}}
    \rho_i\tilde{\boldsymbol{z}}_i
    \right).
    \label{eq:moe_fusion_quamoe}
\end{align}
The final latent representation is obtained by adding the event-sensing dynamic residual and the network-context residual to the MoE fused feature:
\begin{align}
    \boldsymbol{z}_{c}
    =
    \operatorname{LN}
    \left(
    \boldsymbol{z}_{f}
    +
    \lambda_d P_d(\boldsymbol{d}_{es})
    +
    \lambda_n P_n(\boldsymbol{c}_{\rm net})
    \right).
    \label{eq:final_latent_quamoe}
\end{align}
Here $\boldsymbol{c}_{\rm net}$ denotes the network-context representation, and $P_d(\cdot)$ and $P_n(\cdot)$ are projection layers. This formulation follows the implementation in which network context is added as a residual after MoE fusion, rather than being directly routed inside the MoE module.

Quality-aware routing is mainly intended for three failure regimes. When the event map is dropped or sparse, the router should reduce reliance on motion saliency and use geometry, sensing states, and wireless history. When the structured sensing vector is noisy, the calibrated event branch and recent wireless observations become more reliable. When wireless history is incomplete or stale, the model should trust current geometry and sensing more strongly because the future blockage has not yet appeared in past link measurements. The quality vector therefore does not merely provide another feature; it changes the effective estimator used by the fusion block.

\subsection{Communication-Oriented Decision Learning}

QuaMoE-DRF is trained with a joint objective that combines map-level supervision and communication-decision supervision. The map-level losses supervise the normalized beam-independent received-power channel, beam-wise normalized SINR-based link-quality channels, blockage likelihood, and ambiguity. The decision-level losses supervise BS selection, beam selection, joint BS--beam selection, and MCS prediction. When event-sensing dynamic fusion is enabled, an auxiliary blocker-state prediction loss is also included to regularize the dynamic representation.

The map loss is
\begin{align}
    \mathcal{L}_{\rm map}
    &=
    \left\|\hat{\mathbf{Y}}^{\beta}-\mathbf{Y}^{\beta}\right\|_1
    +
    \frac{1}{K}
    \sum_{k=1}^{K}
    \left\|\hat{\mathbf{Y}}^{k}-\mathbf{Y}^{k}\right\|_1,
    \label{eq:map_loss}
\end{align}
and the blockage and ambiguity losses are
\begin{align}
    \mathcal{L}_{\rm blk}
    &=
    \operatorname{BCE}(\hat{\mathbf{Y}}^{\rm blk},\mathbf{Y}^{\rm blk}), \\
    \mathcal{L}_{\rm unc}
    &=
    \left\|\hat{\mathbf{Y}}^{\rm unc}-\mathbf{Y}^{\rm unc}\right\|_2^2.
    \label{eq:blk_unc_losses}
\end{align}
The decision losses are standard cross-entropy losses:
\begin{align}
    \mathcal{L}_{\rm dec}
    &=
    \operatorname{CE}(\boldsymbol{\ell}_{\rm BS},m^{\star})
    +
    \operatorname{CE}(\boldsymbol{\ell}_{\rm beam},k^{\star})
    \notag\\
    &\quad+
    \operatorname{CE}(\boldsymbol{\ell}_{\rm joint},j^{\star})
    +
    \operatorname{CE}(\boldsymbol{\ell}_{\rm MCS},c^{\star}).
    \label{eq:decision_losses}
\end{align}
where $j^{\star}$ indexes the oracle BS--beam pair. The default training objective is written as
\begin{align}
    \mathcal{L}_{\rm def}
    =
    \sum_{q\in\mathcal{D}}
    \lambda_q \mathcal{L}_q,
    \label{eq:default_training_objective_quamoe}
\end{align}
where 
$\mathcal{D}=\{\mathrm{map},\mathrm{blk},\mathrm{unc},\mathrm{dec},\mathrm{aux}\}$.
The auxiliary loss $\mathcal{L}_{\rm aux}$ is active when the dynamic calibration branch predicts blocker-related states. The implementation also supports two optional decision-regularization terms, including a SINR-derived soft beam target and a beam-logit margin loss:
\begin{align}
    \mathcal{L}
    =
    \mathcal{L}_{\rm def}
    +
    \lambda_{\rm soft}\mathcal{L}_{\rm softbeam}
    +
    \lambda_{\rm mar}\mathcal{L}_{\rm margin}.
    \label{eq:full_training_objective_quamoe}
\end{align}
In the main reported configuration, $\lambda_{\rm soft}=0$ and $\lambda_{\rm mar}=0$, and these terms are disabled unless explicitly enabled in tuned variants. This coupled objective discourages the predictor from optimizing pixel-wise map accuracy in isolation and forces the learned dynamic radio map to carry the information required for proactive communication control.

\section{Experimental Setup}
\label{sec:setup}

\subsection{Dataset and Simulation Configuration}

The evaluation is conducted on a dynamic urban ISAC benchmark with multiple BSs, multiple UEs, static road geometry, moving blockers, event-like motion observations, structured sensing states, and wireless history measurements. Each sample contains current multimodal observations and a future local dynamic radio map target. The target definition follows Section~\ref{sec:system_model}.

The benchmark contains $13200$ labeled samples with a fixed $9240/1320/2640$ train/validation/test split. The scene proportions are $33.3\%$ road canyon and $16.7\%$ each for crossroad, T junction, offset crossroad, and complex road; hence the test split contains $880$ canyon samples and $440$ samples from each of the four intersection-like templates. The severe-blockage subset used later is condition-based and overlaps with the layout subsets, rather than forming an additional disjoint partition. The data split uses random seed $2025$, and the neural experiments are repeated with three training seeds $\{2023,2024,2025\}$. The tables report the mean over these runs, with the same ranking observed across seeds.

All compared methods use the same fixed split. Model selection and hyperparameter tuning are performed on the validation split, while the test split is used only for final reporting. The main labels are generated by the compact blockage/path-loss simulator in Section~\ref{sec:system_model}. Sionna RT is used only for a $600$-link calibration and sanity-check subset and is not used to generate the reported training, validation, or test labels. Therefore, the current numerical results demonstrate controlled dynamic-blockage performance, but they should not be interpreted as large-scale ray-tracing validation. The benchmark generator, configuration files, split indices, and evaluation scripts will be released upon publication. As an additional scene-disjoint diagnostic, training on the four non-complex layouts and testing only on the complex-road layout gives $386.9$ Mbps effective rate and $0.0647$ outage for QuaMoE-DRF, indicating degradation under an unseen layout but preserving the same qualitative advantage. The main simulation and learning parameters are summarized in Tables~\ref{tab:setup_simulation}.

% Revision note for the authors: Before final TWC submission, replace the current Sionna sanity-check statement by a large-scale Sionna RT, 3GPP UMi, or DeepSense-like validation table if those experiments are actually run. Do not insert numerical values that have not been produced by the scripts.

\begin{table*}[!t]
\centering
\caption{Main Simulation Parameters}
\label{tab:setup_simulation}
\scriptsize
\resizebox{\textwidth}{!}{
\begin{tabular}{lclc}
\toprule
\textbf{Item} & \textbf{Value} & \textbf{Item} & \textbf{Value} \\
\midrule
Scene types &
Road canyon, crossroad, T junction, offset crossroad, complex road &
Scene size &
$96$ m $\times$ $96$ m \\
Global geometry grid &
$64\times64$ &
Local forecast grid &
$32\times32$ \\
Local forecast extent &
$36$ m $\times$ $36$ m &
History length &
$6$ slots \\
Prediction horizon &
$2$ slots &
Slot duration &
$1$ s \\
Candidate BSs per scene &
$2$--$4$ &
Users per scene &
$3$--$8$ \\
Scene blockers &
$3$--$7$ &
Maximum encoded blockers &
$3$ \\
Candidate beams &
$8$ &
Beam width &
$42^\circ$ \\
Transmit power &
$28$ dBm &
Reference path loss &
$34$ dB at $1$ m \\
Path loss exponent &
$2.45$ &
Dynamic blockage penalty &
$18$ dB \\
Static occlusion penalty &
$12$ dB &
Noise power &
$-94$ dBm \\
Interference power &
$-103$ dBm &
MCS thresholds &
$[-5,-1,3,7,11,15]$ dB \\
MCS spectral efficiencies &
$[0.50,1.00,1.60,2.40,3.20,4.20]$ &
Rate normalization &
$100$ Mbps per b/s/Hz \\
\bottomrule
\end{tabular}}
\end{table*}

\subsection{Compared Methods}

The proposed QuaMoE-DRF is compared with three representative completed baselines. All methods are evaluated on the same test split with the same evaluation protocol.

CKM-UNet is a CKM-style U-Net baseline inspired by RadioUNet~\cite{levie2021radiounet}. It uses static environmental semantics and sparse current wireless observations, with the sparse wireless observation ratio set to $0.10$.

Radar-EKF BF is a sensing-state-based predictive beamforming baseline following the radar-assisted predictive beamforming principle~\cite{liu2020radarassisted}. It extrapolates UE and blocker states from the current sensing vector and then selects the BS, beam, and MCS according to the predicted geometry.

RM+RP is a static radio map and reduced-probing baseline motivated by radio-map-assisted beamforming with reduced pilots~\cite{yang2025reducedpilots}. The main comparison reports RM+RP with $N=2$ probed beams, since this setting gives the best effective rate among the tested probing budgets after overhead accounting.

For RM+RP, the effective rate includes online probing overhead. With $K=8$ candidate beams and a full sweep overhead factor of $0.10$, the overhead ratio for $N=2$ is $0.10\times2/8=0.025$. QuaMoE-DRF, CKM-UNet, and Radar-EKF BF do not use explicit online beam probing in the reported setting.

% Revision note for the authors: The following baselines are strongly recommended before final submission if implementation time allows: (i) Direct-BP-CNN/Transformer using all modalities but no map target, (ii) multimodal fusion without dynamic map supervision, (iii) dynamic CKM-GRU or ConvLSTM map forecasting, (iv) uncertainty-aware fusion without MoE, and (v) camera/LiDAR/semantic beam prediction baselines if compatible data are generated. Keep them commented until actual numerical results are available.

\subsection{Evaluation Metrics}

The primary metric is the effective realized rate, defined as
\begin{align}
    R_{\rm eff}
    =
    (1-\rho_{\rm oh})R_{\rm raw},
\end{align}
where $R_{\rm raw}$ is the realized rate before overhead accounting and $\rho_{\rm oh}$ is the online probing overhead ratio. The realized rate is computed according to the MCS threshold rule in Section~\ref{sec:system_model}. A sample is counted as outage according to \eqref{eq:outage_indicator}.

Communication performance is evaluated by effective rate, raw rate, spectral efficiency, outage probability, rate ratio, BS accuracy, beam accuracy, joint BS--beam accuracy, Beam@3, tied beam accuracy, and MCS accuracy. Beam@3 regards a prediction as correct when the oracle beam is within the top three predicted beam logits for the predicted BS. Tied beam accuracy regards a prediction as correct when the predicted BS is correct and the selected beam belongs to the set of beams that achieve the oracle equivalent MCS-constrained rate. Beam@3 can become high when the beam codebook is small and neighboring beams are communication-equivalent; therefore, Beam@1, tied beam accuracy, outage, and effective rate are reported together.

Map and blockage prediction are evaluated by RMSE, NMSE, MAE, blockage accuracy, blockage F1 score, and blockage transition accuracy. Runtime and complexity are measured by per-sample inference latency and the number of trainable parameters. The saved evaluation files record these communication, map, blockage, overhead, latency, and model size metrics for reproducibility.

\subsection{Hardware and Software Environment}

All experiments were conducted on a workstation with an Intel Core i9-12900KF CPU, an NVIDIA GeForce RTX 5060 Ti GPU with 16 GB memory, and 64 GB RAM. The benchmark generation pipeline was implemented in Python 3.11. The main data-generation path uses the compact blockage/path-loss simulator described in Section~\ref{sec:system_model}; Sionna RT with TensorFlow is retained as an optional calibration backend for the sampled sanity subset only. The forecasting model, baselines, training scripts, and evaluation scripts were implemented in PyTorch 2.x with CUDA 11.8 and cuDNN. Inference latency was measured with CUDA synchronization and reported as the average per-sample runtime on the test split.

\begin{figure}[!t]
\centering
\includegraphics[width=\columnwidth]{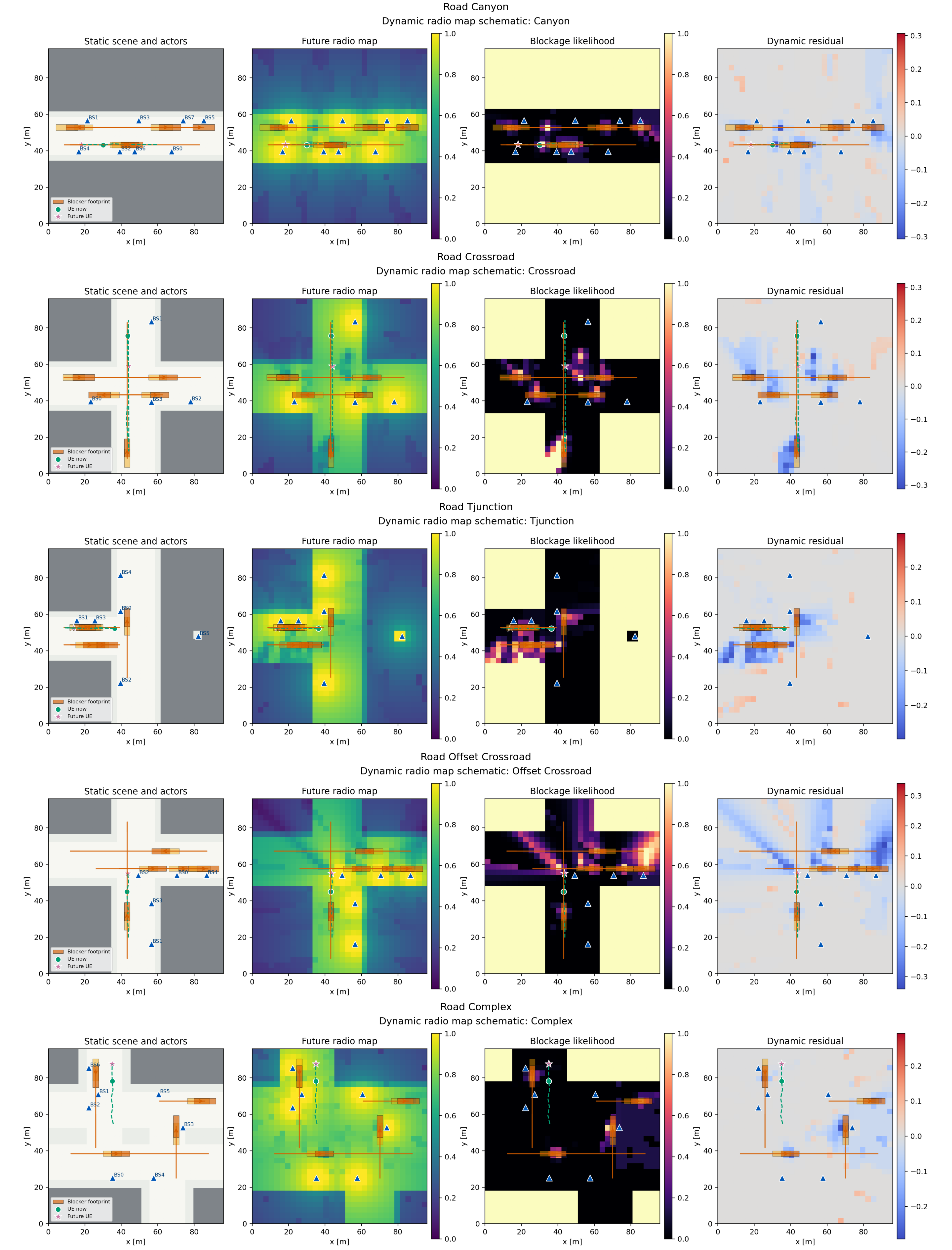}
\caption{Representative dynamic radio map scenarios used in the benchmark. The examples include road canyon, crossroad, T junction, offset crossroad, and complex road layouts. Each case shows the static scene and actors, future $\beta$ map, blockage likelihood, and dynamic residual.}
\label{fig:dynamic_schematics}
\end{figure}

\begin{figure}[!t]
\centering
\includegraphics[width=\columnwidth]{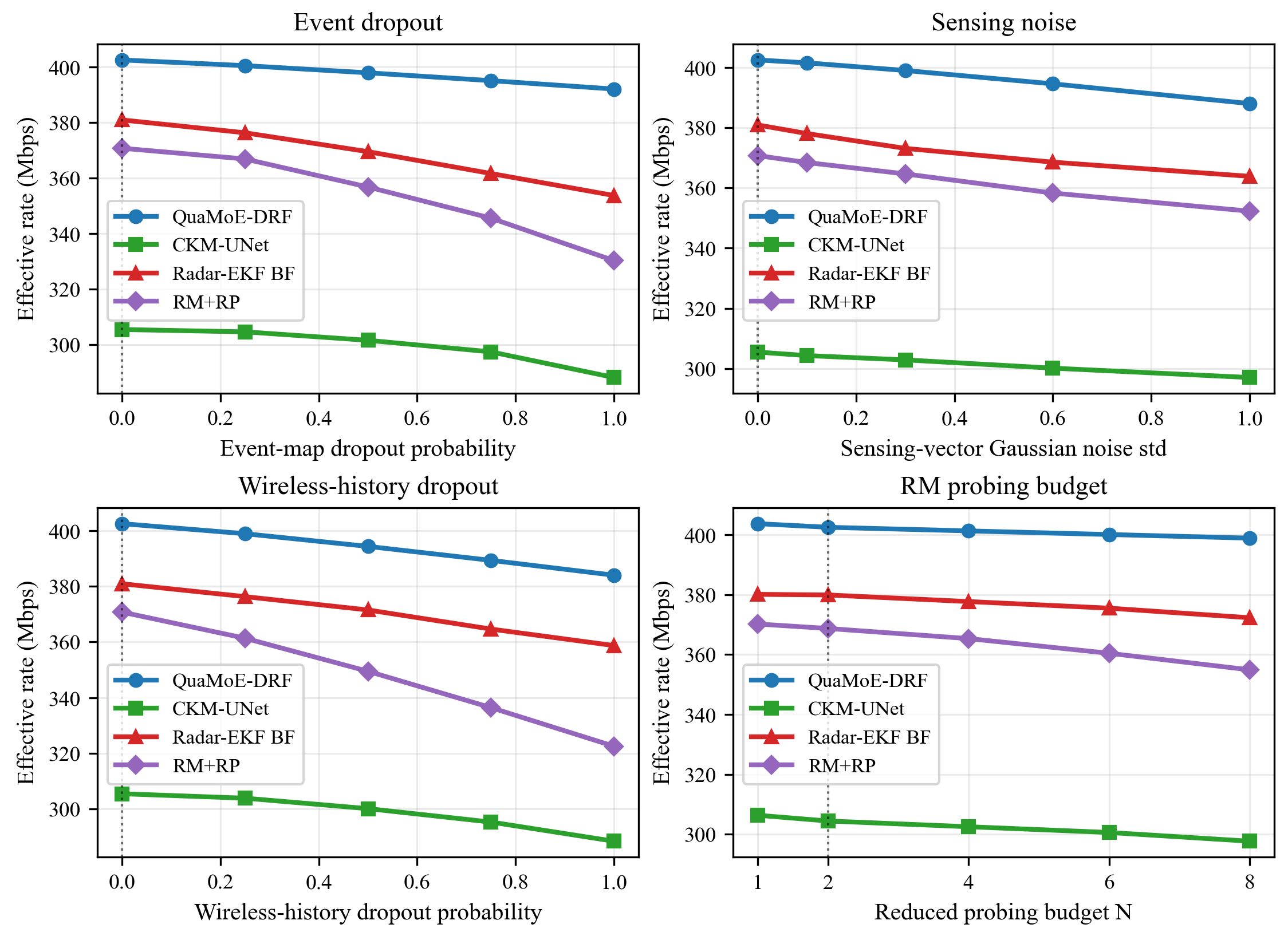}
\caption{Effective rate sensitivity under event map dropout, sensing vector Gaussian noise, wireless history dropout, and reduced-probing budget.}
\label{fig:rate_sensitivity}
\end{figure}

\section{Results and Analysis}
\label{sec:results}

\subsection{Dynamic Radio Map Visualization}

Fig.~\ref{fig:dynamic_schematics} illustrates representative dynamic radio map samples in the benchmark. The selected scenarios cover road canyon, crossroad, T junction, offset crossroad, and complex road layouts, which contain different BS placements, moving blockers, and future UE positions. The future $\beta$ maps show that the communication-oriented radio map varies smoothly in open road regions but changes rapidly near blocker footprints and street intersections. The blockage likelihood maps further reveal that the dominant degradation is spatially concentrated around the dynamic blockers and the possible future UE trajectory. The dynamic residual maps show that the largest map variation appears near the blocker-induced shadowing regions, while distant regions remain relatively stable.

The visualization makes the role of a dynamic radio map explicit. A static channel knowledge map cannot directly describe the future obstruction state caused by mobile blockers. In contrast, the proposed predictor uses multimodal sensing and wireless history to infer the future dynamic radio map, the blockage pattern, and the residual dynamic change. This provides a more suitable basis for proactive BS association, beam selection, and rate adaptation.

\subsection{Overall Communication Performance}

Table~\ref{tab:core_comm_performance} reports the final communication performance of the proposed QuaMoE-DRF and the three representative completed baselines on the fixed test split. For RM+RP, the $N=2$ setting is reported in the main comparison because it provides the best effective rate among the tested probing settings after considering probing overhead.

QuaMoE-DRF achieves the best effective rate, outage probability, Beam@3 accuracy, and tied beam accuracy among the compared methods. Compared with the strongest effective-rate baseline, Radar-EKF BF, it improves the effective rate by $5.67\%$ and reduces the outage probability by $8.35\%$. The absolute gain is moderate rather than overwhelming, so the result should be interpreted together with the theoretical margin analysis, map reconstruction results, ablation study, and stress tests. Compared with the strongest reduced-probing baseline, RM+RP $N=2$, the gain is larger because QuaMoE-DRF avoids explicit online beam probing while exploiting multimodal prediction. This result is consistent with the intended use of the forecast map: it replaces part of the online search by a prediction of the future beam-SINR field.

\begin{table}[htbp]
\centering
\caption{Final Test Communication Performance of the Core Methods}
\label{tab:core_comm_performance}
\scriptsize
\resizebox{\columnwidth}{!}{
\begin{tabular}{lccccccc}
\toprule
\textbf{Method} &
\textbf{Eff. Rate} &
\textbf{SE} &
\textbf{Outage} &
\textbf{Rate Ratio} &
\textbf{Beam@1} &
\textbf{Beam@3} &
\textbf{Tied Beam} \\
& \textbf{(Mbps)} & & & & & & \\
\midrule
\textbf{QuaMoE-DRF}
& \textbf{402.5} & \textbf{4.025}
& \textbf{0.0417} & \textbf{0.9583}
& \textbf{0.8836}
& \textbf{1.0000} & \textbf{0.9773} \\

CKM-UNet
& 305.5 & 3.055
& 0.2727 & 0.7273
& 0.7235
& 0.8173 & 0.7691 \\

Radar-EKF BF
& 380.9 & 3.809
& 0.0455 & 0.9069
& 0.8439
& 0.9561 & 0.8897 \\

RM+RP $N=2$
& 370.7 & 3.707
& 0.0827 & 0.8826
& 0.8182
& 0.9014 & 0.8467 \\
\bottomrule
\end{tabular}}
\end{table}

The Beam@3 value of $1.0000$ does not mean that the beam task is trivial. The candidate codebook has only eight beams, and neighboring beams can support the same MCS under the threshold-rate rule. This is why strict Beam@1 is $0.8836$ while tied beam accuracy is $0.9773$. Effective rate and outage remain the primary communication metrics.

\subsection{Map Prediction and Blockage Estimation}

Table~\ref{tab:core_map_performance} reports the map-level and blockage-level metrics of the compared methods. QuaMoE-DRF achieves the best map and blockage prediction performance, indicating that the communication gain is supported by improved future dynamic radio map reconstruction rather than by downstream decision tuning alone. Radar-EKF BF obtains competitive communication performance in Table~\ref{tab:core_comm_performance}, but its weaker map-level metrics show that tracking-based prediction lacks an explicit communication-oriented map representation. Overall, the results indicate that improving the predicted future radio map also improves the reliability of the downstream transmission decision.

\begin{table}[!t]
\centering
\caption{Final Test Map and Blockage Prediction Metrics of the Core Methods}
\label{tab:core_map_performance}
\scriptsize
\resizebox{\columnwidth}{!}{
\begin{tabular}{lccccc}
\toprule
\textbf{Method} &
\textbf{RMSE} &
\textbf{NMSE} &
\textbf{MAE} &
\textbf{Blockage Acc.} &
\textbf{Blockage F1} \\
\midrule
\textbf{QuaMoE-DRF}
& \textbf{0.1836} & \textbf{0.0620} & \textbf{0.1356}
& \textbf{0.9645} & \textbf{0.9280} \\

CKM-UNet
& 0.2215 & 0.0855 & 0.1685
& 0.9255 & 0.8885 \\

Radar-EKF BF
& 0.3124 & 0.1012 & 0.2037
& 0.9115 & 0.8716 \\

RM+RP $N=2$
& 0.2174 & 0.0821 & 0.1670
& 0.9262 & 0.8878 \\
\bottomrule
\end{tabular}}
\end{table}

\subsection{Robustness and Sensitivity Analysis}

Fig.~\ref{fig:rate_sensitivity} further evaluates the robustness of the compared methods under four stress factors. Under event map dropout, the effective rate of QuaMoE-DRF decreases only from approximately $402$ Mbps to approximately $392$ Mbps as the dropout probability increases to $1.0$. In contrast, RM+RP drops from approximately $370$ Mbps to approximately $331$ Mbps, and Radar-EKF BF drops from approximately $381$ Mbps to approximately $354$ Mbps. CKM-UNet remains the weakest method over the whole range. This shows that the proposed method does not rely on the event map alone, because the wireless history and sensing features can still support future dynamic radio map inference when event information is partially missing.

Under sensing vector Gaussian noise, QuaMoE-DRF also maintains the highest effective rate. When the noise standard deviation increases to $1.0$, the proposed method still achieves approximately $388$ Mbps. Radar-EKF BF, RM+RP, and CKM-UNet decrease to approximately $364$ Mbps, $352$ Mbps, and $297$ Mbps, respectively. This result indicates that the quality-aware fusion mechanism can reduce the effect of noisy sensing features and preserve stable communication decisions.

The wireless history dropout curve provides the most direct evidence for multimodal robustness. As the dropout probability increases from $0$ to $1.0$, QuaMoE-DRF decreases from approximately $403$ Mbps to approximately $384$ Mbps. The degradation is smaller than that of RM+RP, which falls to approximately $323$ Mbps, and is also smaller than that of Radar-EKF BF. This behavior indicates that the proposed method falls back on complementary sensing and event information when historical wireless observations become incomplete.

The reduced-probing budget curve also shows the advantage of prediction-based transmission. QuaMoE-DRF remains close to $400$ Mbps over the tested range, while RM+RP decreases as the effective probing resource becomes more constrained. Since RM+RP depends on online probing to compensate for incomplete radio map information, its effective rate is more sensitive to the probing setting. In contrast, QuaMoE-DRF directly predicts the future decision map from available multimodal information and avoids the explicit probing overhead. Therefore, Fig.~\ref{fig:rate_sensitivity} shows that the proposed method is not only strong under the nominal setting, but also robust to missing events, noisy sensing, incomplete wireless history, and limited probing resources. These stress tests also clarify when quality-aware routing is needed. The advantage is most visible when the reliability pattern changes at test time: event dropout, sensing noise, and wireless-history dropout each degrade a different modality, but the full model keeps the effective-rate loss within a moderate range by shifting attention to the remaining reliable sources. This behavior supports the theoretical view that fixed fusion is suboptimal under heteroscedastic modality uncertainty.

\subsection{Subset and Tied Beam Analysis of the Proposed Method}

Table~\ref{tab:moe_subset_tied} reports the subset-level performance of QuaMoE-DRF. The layout rows are defined from the scene template: street canyon corresponds to the road-canyon layout, while the intersection row pools crossroad, T junction, offset-crossroad, and complex-road layouts. The severe blockage row is a condition-defined subset and overlaps with these layout rows, which explains why the subset counts are not intended to sum to the overall test size. The overall effective rate is $402.5$ Mbps with an outage probability of $0.0417$. In the severe blockage subset, the effective rate decreases to $394.3$ Mbps and the outage probability increases to $0.0611$, which shows that deep blockage remains the most challenging condition. Nevertheless, the tied beam accuracy remains $0.9667$, indicating that the selected beams are still close to the communication optimal choices in most samples.

The street canyon subset gives the best result, with an effective rate of $415.2$ Mbps and an outage probability of $0.0114$. This is consistent with the more regular propagation structure in canyon-like road layouts. The intersection subset has an effective rate of $396.1$ Mbps and an outage probability of $0.0568$, which reflects the stronger uncertainty caused by crossing blockers and more frequent link transitions.

The gap between strict Beam@1 accuracy and tied beam accuracy is also important. The overall Beam@1 accuracy is $0.8836$, while the tied beam accuracy reaches $0.9773$. This means that many strict beam index mismatches still lead to communication-equivalent decisions. In practical beam codebook systems, neighboring beams may provide similar SINR and support the same MCS. Therefore, effective rate, outage probability, Beam@3 accuracy, and tied beam accuracy are more informative than Beam@1 accuracy alone.

\begin{table}[!t]
\centering
\caption{Subset and Tied Beam Analysis of the Proposed QuaMoE-DRF}
\label{tab:moe_subset_tied}
\scriptsize
\resizebox{\columnwidth}{!}{
\begin{tabular}{lccccc}
\toprule
\textbf{Subset} &
\textbf{Samples} &
\textbf{Rate} &
\textbf{Outage} &
\textbf{Beam@1} &
\textbf{Tied Beam} \\
& & \textbf{(Mbps)} & & & \\
\midrule
Overall & 2640 & 402.5 & 0.0417 & 0.8836 & 0.9773 \\
Severe blockage & 1800 & 394.3 & 0.0611 & 0.8000 & 0.9667 \\
Street canyon & 880 & 415.2 & 0.0114 & 0.9886 & 0.9886 \\
Intersection & 1760 & 396.1 & 0.0568 & 0.8011 & 0.9716 \\
\bottomrule
\end{tabular}}
\end{table}

\subsection{Ablation Study}

To further verify the contribution of the main components, Table~\ref{tab:ablation_study} reports the ablation results of QuaMoE-DRF. The full model achieves the best overall communication and map prediction performance. Replacing MoE fusion with direct feature concatenation reduces the effective rate from $402.5$ Mbps to $390.8$ Mbps and raises outage from $0.0417$ to $0.0568$. Removing the quality vector gives $393.6$ Mbps and $0.0530$ outage. These two rows are important for the novelty claim: the gain is not from simply stacking several modules, but from allowing the estimator to change with sample-dependent modality reliability. In relative terms, discarding the quality vector increases outage by about $27\%$, which is consistent with the heteroscedastic-fusion analysis in Theorem~3.

The single-modality removal results further show that all modalities contribute to the final performance. Removing static geometry weakens the spatial propagation prior, while removing event-like motion or structured sensing states reduces the ability to capture dynamic blockage evolution. Removing wireless history also causes a clear performance loss, indicating that recent link observations provide useful communication evidence complementary to environmental sensing. In addition, the map-only variant obtains competitive map reconstruction accuracy but weaker communication decisions, whereas the decision-only variant achieves reasonable communication performance but suffers from degraded map prediction. This confirms that joint map-and-decision learning is necessary to align dynamic radio map prediction with proactive BS, beam, and MCS adaptation.

\begin{table}[!t]
\centering
\caption{Ablation Study of the Proposed QuaMoE-DRF}
\label{tab:ablation_study}
\scriptsize
\resizebox{\columnwidth}{!}{
\begin{tabular}{lcccccc}
\toprule
\textbf{Variant} &
\textbf{Eff. Rate} &
\textbf{Outage} &
\textbf{Beam@3} &
\textbf{Tied Beam} &
\textbf{Map RMSE} &
\textbf{Blockage F1} \\
& \textbf{(Mbps)} & & & & & \\
\midrule
\textbf{Full QuaMoE-DRF}
& \textbf{402.5} & \textbf{0.0417} & \textbf{1.0000} & \textbf{0.9773} & \textbf{0.1836} & \textbf{0.9280} \\

w/o MoE fusion
& 390.8 & 0.0568 & 0.9826 & 0.9424 & 0.1947 & 0.9145 \\

w/o quality vector
& 393.6 & 0.0530 & 0.9879 & 0.9515 & 0.1912 & 0.9184 \\

w/o static geometry
& 385.4 & 0.0682 & 0.9659 & 0.9258 & 0.2079 & 0.9021 \\

w/o event-like motion
& 392.2 & 0.0576 & 0.9818 & 0.9477 & 0.1964 & 0.9106 \\

w/o sensing state
& 386.7 & 0.0648 & 0.9735 & 0.9330 & 0.2038 & 0.9047 \\

w/o wireless history
& 388.1 & 0.0621 & 0.9773 & 0.9364 & 0.2005 & 0.9089 \\

Map-only training
& 371.3 & 0.0909 & 0.9205 & 0.8614 & 0.1868 & 0.9241 \\

Decision-only training
& 382.4 & 0.0723 & 0.9545 & 0.9121 & 0.2458 & 0.8720 \\
\bottomrule
\end{tabular}}
\end{table}

\section{Conclusion}
\label{sec:conclusion}

This paper presented QuaMoE-DRF, a communication-oriented dynamic radio map forecasting framework for proactive beam and rate adaptation in ISAC networks. The key representation is the future beam-SINR field: the full multi-BS field is sufficient for finite-codebook threshold-rate decisions, while the compact reference-BS projection used in implementation is not sufficient for BS association alone. Accordingly, QuaMoE-DRF combines this compact local field with BS-level supervision, joint BS--beam supervision, and latent network context. It further fuses static geometry, event-like motion, structured sensing states, and wireless history through quality-aware mixture-of-experts routing, so that the predictor can adapt to modality-dependent reliability changes. The joint map-and-decision objective aligns dynamic field prediction with BS association, beam selection, and MCS adaptation. Simulations on a dynamic urban benchmark show improved effective rate, outage probability, beam robustness, blockage estimation, and map accuracy over representative baselines. Future work will strengthen external validation using larger ray-tracing or real-world multimodal datasets and broader direct beam-prediction baselines.

\appendices

\section{Proofs and Communication-Oriented Error Bounds}
\label{app:proofs}

\begingroup
\setlength{\jot}{2pt}

This appendix connects the representation and fusion design of QuaMoE-DRF to communication-level reliability. The derivation is organized as a sequence of linked results: decision sufficiency of the beam-SINR field, non-sufficiency of the compact projection, CRLB-induced SINR-map error, inverse-uncertainty fusion, margin-based outage and MCS backoff, and reduced-probing effective-rate loss. CRLB-type sensing metrics are widely used in ISAC performance analysis \cite{LiuTSP2022CRB,HuaTWC2024CRBRate,RenTWC2024CRBRate}. The future-link prediction viewpoint follows sensing-assisted predictive beamforming \cite{liu2020radarassisted}, while the reduced-probing interpretation is consistent with radio-map-assisted beamforming with reduced pilots \cite{yang2025reducedpilots}.

\subsection{Proof of Theorem~1}

Consider one future grid point $\boldsymbol{x}\in\Omega_{t+\Delta}$. Define the candidate link set, the link SINR, and the full candidate beam-SINR field as
\begin{subequations}
\label{eq:app_decision_state}
\begin{align}
    \mathcal{L}
    &\triangleq
    \mathcal{B}\times\{1,\ldots,K\}, \\
    \ell
    &\triangleq
    (m,k)\in\mathcal{L}, \\
    \gamma_{\ell}
    &\triangleq
    \gamma^{\rm dB}_{m,k}(\boldsymbol{x},t+\Delta), \\
    \boldsymbol{\gamma}_{\boldsymbol{x}}
    &\triangleq
    \{\gamma_{\ell}:\ell\in\mathcal{L}\}.
\end{align}
\end{subequations}
Under the threshold-MCS rule in \eqref{eq:mcs_index}--\eqref{eq:oracle_decision}, the MCS index, raw rate, and oracle BS--beam link are deterministic functions of $\boldsymbol{\gamma}_{\boldsymbol{x}}$:
\begin{subequations}
\label{eq:app_decision_rule}
\begin{align}
    c_{\ell}^{\star}
    &=
    \max
    \Big(
    \{-1\}
    \cup
    \{c\in\mathcal{C}:\gamma_{\ell}\ge\Gamma_c\}
    \Big), \\
    R_{\ell}^{\star}
    &=
    B_0\eta_{c_{\ell}^{\star}}, \\
    \ell^{\star}
    &=
    \underset{\ell\in\mathcal{L}}{\arg\max}\;
    \big(R_{\ell}^{\star},\gamma_{\ell}\big), \\
    (m^{\star},k^{\star})
    &=
    \ell^{\star}.
\end{align}
\end{subequations}
The selected MCS is $c_{\ell^{\star}}^{\star}$, the realized oracle rate is $R_{\ell^{\star}}^{\star}$, and the outage indicator follows directly from \eqref{eq:outage_indicator}. The tied communication-equivalent beam set for the selected BS is
\begin{equation}
    \mathcal{T}_{m^{\star}}
    =
    \{k: B_0\eta_{c_{m^{\star},k}^{\star}}=R_{\ell^{\star}}^{\star}\}.
\end{equation}
It is also a deterministic function of $\boldsymbol{\gamma}_{\boldsymbol{x}}$. Therefore, for the finite-codebook threshold-rate model, the full multi-BS beam-SINR field is decision-sufficient. \hfill$\blacksquare$

\subsection{Proof of Proposition~1}

It suffices to give a counterexample. Let there be two BSs, one beam per BS, and a single future point. Consider two full fields
\begin{align}
    \boldsymbol{\gamma}^{(A)}=(10,9),
    \quad
    \boldsymbol{\gamma}^{(B)}=(9,10),
\end{align}
where the two entries correspond to BS~1 and BS~2. Under any strictly increasing tie-breaking rule based on SINR or threshold rate, the oracle serving BS is BS~1 for field $A$ and BS~2 for field $B$. If the compact projection stores only the reference serving SINR slice and does not include the BS identity, then both projected maps equal the scalar value $10$. Hence the same compact projection corresponds to different oracle BS decisions. Therefore, the projection alone is not sufficient for multi-BS association. In QuaMoE-DRF, the missing information is addressed by BS logits, joint BS--beam logits, and latent network context, not by claiming that the projected map alone is sufficient. \hfill$\blacksquare$

\subsection{CRLB-Limited SINR Forecasting Error}

Let $\boldsymbol{\xi}_{t+\Delta}$ denote the future dynamic state that affects the link, including the UE position and relevant blocker states. The local link model and the modality-dependent observation model are written as
\begin{subequations}
\label{eq:app_observation_group}
\begin{align}
    \gamma_{\ell}
    &=
    G_{\ell}(\boldsymbol{\xi}_{t+\Delta}), \\
    \mathcal{M}_{\rm mod}
    &\triangleq
    \{g,e,s,w\}, \\
    \mathbf{s}_i
    &\triangleq
    \nabla_{\boldsymbol{\xi}_{t}}
    \log p_i(\mathbf y_i|\boldsymbol{\xi}_{t},q_i), \\
    \mathbf{J}_i(q_i)
    &=
    \mathbb{E}\big[\mathbf{s}_i\mathbf{s}_i^{\mathsf T}\big],
    \quad
    i\in\mathcal{M}_{\rm mod}.
\end{align}
\end{subequations}
The future-state prediction and its propagated CRLB are grouped as
\begin{subequations}
\label{eq:app_crlb_group}
\begin{align}
    \boldsymbol{\xi}_{t+\Delta}
    &=
    \mathbf{F}_{\Delta}\boldsymbol{\xi}_{t}
    +
    \boldsymbol{\omega}_{\Delta}, \\
    \mathbb{E}[
    \boldsymbol{\omega}_{\Delta}
    \boldsymbol{\omega}_{\Delta}^{\mathsf T}]
    &=
    \mathbf{Q}_{\Delta}, \\
    \mathbf{P}_{i}^{\rm obs}(\Delta)
    &\triangleq
    \mathbf{F}_{\Delta}
    \mathbf{J}_{i}^{-1}(q_i)
    \mathbf{F}_{\Delta}^{\mathsf T}, \\
    \mathbf{P}_{i}(\Delta)
    &\triangleq
    \mathbf{P}_{i}^{\rm obs}(\Delta)
    +
    \mathbf{Q}_{\Delta}, \\
    {\rm cov}(\hat{\boldsymbol{\xi}}_{i,t+\Delta})
    &\succeq
    \mathbf{P}_{i}(\Delta).
\end{align}
\end{subequations}
The state-estimation error is then propagated to the dB-domain SINR prediction error by first-order linearization:
\begin{subequations}
\label{eq:app_sinr_error_group}
\begin{align}
    \mathbf{h}_{\ell}
    &\triangleq
    \nabla_{\boldsymbol{\xi}}G_{\ell}(\boldsymbol{\xi})
    \big|_{\boldsymbol{\xi}=\boldsymbol{\xi}_{t+\Delta}}, \\
    \sigma_{i,\ell}^{2}
    &\triangleq
    {\rm var}
    \big(
    \hat\gamma_{\ell}^{(i)}
    -
    \gamma_{\ell}
    \big|q_i
    \big), \\
    \underline{\sigma}_{i,\ell}^{2}
    &\triangleq
    \mathbf{h}_{\ell}^{\mathsf T}
    \mathbf{P}_{i}(\Delta)
    \mathbf{h}_{\ell}, \\
    \sigma_{i,\ell}^{2}
    &\ge
    \underline{\sigma}_{i,\ell}^{2}.
\end{align}
\end{subequations}
Equations \eqref{eq:app_crlb_group} and \eqref{eq:app_sinr_error_group} show that the irreducible SINR-map error increases when the sensing Fisher information decreases, the prediction horizon increases, or the blocker process becomes more uncertain.

\subsection{Local Link Sensitivity}

To identify the regions where the CRLB-limited state error is most harmful, suppose that the UE-position component is $\boldsymbol{x}$ and $d_m(\boldsymbol{x})>d_0$. From \eqref{eq:large_scale_power}--\eqref{eq:beam_sinr}, the position sensitivity of $G_{\ell}$ can be decomposed into path-loss, beam-pattern, and blockage terms:
\begin{subequations}
\label{eq:app_gradient_group}
\begin{align}
    \nabla_{\boldsymbol{x}}G_{\ell}
    &=
    \mathbf{h}_{\ell}^{\rm pl}
    +
    \mathbf{h}_{\ell}^{\rm bm}
    +
    \mathbf{h}_{\ell}^{\rm blk}, \\
    \mathbf{h}_{\ell}^{\rm pl}
    &=
    -\frac{10\alpha}{\ln 10}
    \frac{\boldsymbol{x}-\boldsymbol{b}_m}
    {d_m^2(\boldsymbol{x})}, \\
    \mathbf{r}_{m}^{\perp}(\boldsymbol{x})
    &\triangleq
    \begin{bmatrix}
    -[\boldsymbol{x}-\boldsymbol{b}_m]_2 \\
    [\boldsymbol{x}-\boldsymbol{b}_m]_1
    \end{bmatrix}, \\
    \mathbf{h}_{\ell}^{\rm bm}
    &=
    \frac{
    g_k'(\phi_m(\boldsymbol{x}))
    }
    {d_m^2(\boldsymbol{x})}
    \mathbf{r}_{m}^{\perp}(\boldsymbol{x}), \\
    \mathbf{h}_{\ell}^{\rm blk}
    &=
    \mathbf{h}_{\ell}^{\rm sta}
    +
    \mathbf{h}_{\ell}^{\rm dyn}, \\
    \mathbf{h}_{\ell}^{\rm sta}
    &=
    -\lambda_{\rm sta}
    \nabla_{\boldsymbol{x}}a_m^{\rm sta}(\boldsymbol{x}), \\
    \mathbf{h}_{\ell}^{\rm dyn}
    &=
    -\lambda_{\rm dyn}
    \nabla_{\boldsymbol{x}}a_m^{\rm dyn}(\boldsymbol{x},t+\Delta), \\
    g_k'(\phi)
    &=
    -\frac{
    24\angle(\phi-\bar\phi_k)
    }
    {\Phi_{\rm bw}^{2}}.
\end{align}
\end{subequations}
The last line holds in the unclipped beam region. At clipping points, it can be interpreted through a subgradient or a smoothed beam-gain approximation. Hence the same state-estimation error produces larger SINR uncertainty near sharp beam edges and blockage boundaries.

\subsection{Proof of Theorem~3}

Assume that the modality-specific SINR estimates are locally unbiased and conditionally uncorrelated:
\begin{subequations}
\label{eq:app_unbiased_group}
\begin{align}
    \hat\gamma_{\ell}^{(i)}
    &=
    \gamma_{\ell}
    +
    \varepsilon_{i,\ell}, \\
    \mathbb{E}[\varepsilon_{i,\ell}|q_i]
    &=
    0, \\
    {\rm var}(\varepsilon_{i,\ell}|q_i)
    &=
    \sigma_{i,\ell}^{2}, \\
    {\rm cov}(\varepsilon_{i,\ell},\varepsilon_{j,\ell}|q_i,q_j)
    &=
    0,
    \quad i\ne j.
\end{align}
\end{subequations}
The linear fusion model and the unbiasedness constraint are
\begin{subequations}
\label{eq:app_linear_fusion_group}
\begin{align}
    \hat\gamma_{\ell}
    &=
    \sum_{i\in\mathcal{M}_{\rm mod}}
    \rho_i\hat\gamma_{\ell}^{(i)}, \\
    \sum_{i\in\mathcal{M}_{\rm mod}}
    \rho_i
    &=
    1, \\
    {\rm var}(\hat\gamma_{\ell}-\gamma_{\ell})
    &=
    \sum_{i\in\mathcal{M}_{\rm mod}}
    \rho_i^2\sigma_{i,\ell}^{2}.
\end{align}
\end{subequations}
The conditional MSE minimization is
\begin{subequations}
\label{eq:app_fusion_solution_group}
\begin{align}
    \boldsymbol{\rho}^{\star}
    &=
    \underset{\sum_i\rho_i=1}{\arg\min}
    \sum_{i\in\mathcal{M}_{\rm mod}}
    \rho_i^2\sigma_{i,\ell}^{2}.
\end{align}
\end{subequations}
Using the Lagrangian $\mathcal{J}=\sum_i\rho_i^2\sigma_{i,\ell}^{2}+\lambda(\sum_i\rho_i-1)$ gives $2\rho_i^{\star}\sigma_{i,\ell}^{2}+\lambda^{\star}=0$. Enforcing $\sum_i\rho_i^{\star}=1$ yields
\begin{align}
    \rho_i^{\star}
    &={\sigma_{i,\ell}^{-2}}
    \left/{
    \sum_{j\in\mathcal{M}_{\rm mod}}
    \sigma_{j,\ell}^{-2}}
    \right., \\
    \sigma_{f,\ell}^{2,\star}
    &=
    \left(
    \sum_{j\in\mathcal{M}_{\rm mod}}
    \sigma_{j,\ell}^{-2}
    \right)^{-1}.
\end{align}
Since $\sigma_{i,\ell}^{2}$ is sample-, modality-, and horizon-dependent through \eqref{eq:app_sinr_error_group}, fixed concatenation is generally suboptimal. \hfill$\blacksquare$

\subsection{Proof of Theorem~2 and MCS Backoff}

Let the decision head select $\hat\ell=(\hat m,\hat k)$ and $\hat c$ from the predicted field. Define the prediction error and the MCS margin as
\begin{subequations}
\label{eq:app_margin_group}
\begin{align}
    \delta_{\hat\ell}
    &\triangleq
    \hat\gamma_{\hat\ell}
    -
    \gamma_{\hat\ell}, \\
    \mu_{\hat\ell,\hat c}
    &\triangleq
    \hat\gamma_{\hat\ell}
    -
    \Gamma_{\hat c}, \\
    [z]_+
    &\triangleq
    \max\{z,0\}.
\end{align}
\end{subequations}
The selected-link outage probability satisfies
\begin{subequations}
\label{eq:app_outage_group}
\begin{align}
    P_{\rm out}
    &=
    \Pr\{
    \gamma_{\hat\ell}
    <
    \Gamma_{\hat c}
    \}, \\
    P_{\rm out}
    &=
    \Pr\{
    \delta_{\hat\ell}
    >
    \mu_{\hat\ell,\hat c}
    \}.
\end{align}
\end{subequations}
If $\delta_{\hat\ell}$ is sub-Gaussian with variance proxy $\sigma_{f,\hat\ell}^{2}$, then
\begin{align}
    P_{\rm out}
    &\le
    \exp
    \left(
    -\frac{
    [\mu_{\hat\ell,\hat c}]_+^2
    }
    {2\sigma_{f,\hat\ell}^{2}}
    \right).
\end{align}
Multiplying the success probability by $B_0\eta_{\hat c}$ gives \eqref{eq:main_margin_goodput_bound}. If only the second moment is assumed, Cantelli's inequality gives the distribution-free bound
\begin{subequations}
\label{eq:app_cantelli_group}
\begin{align}
    P_{\rm out}
    &\le
    \frac{
    \sigma_{f,\hat\ell}^{2}
    }
    {
    \sigma_{f,\hat\ell}^{2}
    +
    [\mu_{\hat\ell,\hat c}]_+^2
    }, \\
    \mathbb{E}[R_{\hat\ell,\hat c}]
    &\ge
    B_0\eta_{\hat c}
    \frac{[\mu_{\hat\ell,\hat c}]_+^2}
    {\sigma_{f,\hat\ell}^{2}+[\mu_{\hat\ell,\hat c}]_+^2}.
\end{align}
\end{subequations}
For a target outage level $\epsilon\in(0,1)$, the sub-Gaussian and Cantelli safety backoffs are
\begin{subequations}
\label{eq:app_backoff_group}
\begin{align}
    b_{\epsilon,\ell}^{\rm SG}
    &\triangleq
    \sqrt{
    2\sigma_{f,\ell}^{2}
    \log(1/\epsilon)
    }, \\
    b_{\epsilon,\ell}^{\rm C}
    &\triangleq
    \sigma_{f,\ell}
    \sqrt{
    \frac{1-\epsilon}{\epsilon}
    }, \\
    \mu_{\ell,c}
    &\triangleq
    \hat\gamma_{\ell}
    -
    \Gamma_c.
\end{align}
\end{subequations}
Using the sub-Gaussian backoff, an outage-constrained MCS can be selected as
\begin{subequations}
\label{eq:app_robust_mcs_group}
\begin{align}
    \mathcal{C}_{\epsilon,\ell}^{\rm SG}
    &\triangleq
    \{c\in\mathcal{C}:
    \mu_{\ell,c}
    \ge
    b_{\epsilon,\ell}^{\rm SG}
    \}, \\
    c_{\epsilon}^{\star}(\ell)
    &=
    \max
    \Big(
    \{-1\}
    \cup
    \mathcal{C}_{\epsilon,\ell}^{\rm SG}
    \Big), \\
    \Pr\{
    \gamma_{\ell}
    <
    \Gamma_{c_{\epsilon}^{\star}(\ell)}
    \}
    &\le
    \epsilon, \\
    \mathbb{E}[R_{\ell,c_{\epsilon}^{\star}}]
    &\ge
    (1-\epsilon)
    B_0
    \eta_{c_{\epsilon}^{\star}(\ell)}.
\end{align}
\end{subequations}
Therefore, the same fused uncertainty $\sigma_{f,\ell}^{2}$ used by the quality-aware fusion module also determines how conservative the MCS decision should be. A reliable prediction permits a smaller backoff, whereas an uncertain prediction forces a lower MCS even when the predicted SINR is high.

\subsection{Reduced-Probing Effective-Rate Bound}

The same uncertainty-margin structure also characterizes reduced probing. Let $\mathcal{S}_N$ be the top-$N$ candidate links ranked by predicted SINR:
\begin{subequations}
\label{eq:app_probe_set_group}
\begin{align}
    \mathcal{S}_N
    &=
    \{s_1,\ldots,s_N\}, \\
    \hat\gamma_{s_1}
    &\ge
    \cdots
    \ge
    \hat\gamma_{s_N}, \\
    \hat\gamma_{s_N}
    &\ge
    \hat\gamma_{\ell},
    \quad
    \ell\notin\mathcal{S}_N.
\end{align}
\end{subequations}
For an unprobed link $\ell\notin\mathcal{S}_N$, define its predicted separation from the weakest probed link and the corresponding pairwise overtaking bound as
\begin{subequations}
\label{eq:app_probe_pair_group}
\begin{align}
    d_{\ell,N}
    &\triangleq
    [
    \hat\gamma_{s_N}
    -
    \hat\gamma_{\ell}
    ]_+, \\
    v_{\ell,N}
    &\triangleq
    \sigma_{f,\ell}^{2}
    +
    \sigma_{f,s_N}^{2}, \\
    \upsilon_{\ell,N}
    &\triangleq
    \exp
    \left(
    -\frac{d_{\ell,N}^{2}}
    {2v_{\ell,N}}
    \right), \\
    \Pr\{
    \gamma_{\ell}
    \ge
    \gamma_{s_N}
    \}
    &\le
    \upsilon_{\ell,N}.
\end{align}
\end{subequations}
Applying the union bound gives the miss probability of the probed candidate set:
\begin{subequations}
\label{eq:app_probe_miss_group}
\begin{align}
    p_{\rm miss}(N)
    &\triangleq
    \Pr\{
    \ell^{\star}
    \notin
    \mathcal{S}_N
    \}, \\
    p_{\rm miss}(N)
    &\le
    \sum_{\ell\notin\mathcal{S}_N}
    \upsilon_{\ell,N}, \\
    \bar p_{\rm miss}(N)
    &\triangleq
    \mathbb{E}[p_{\rm miss}(N)].
\end{align}
\end{subequations}
This bound is conservative for threshold-rate communication, because an unprobed link with a larger SINR may still support the same MCS and cause no rate loss. Let $R^{\star}$ be the oracle raw rate over all candidates. The probing-limited raw rate and the probing overhead are
\begin{subequations}
\label{eq:app_probe_rate_group}
\begin{align}
    R_{\max}
    &\triangleq
    B_0\eta_{C-1}, \\
    R_{\rm raw}^{(N)}
    &\ge
    R^{\star}
    -
    R_{\max}
    \mathbf{1}
    \{
    \ell^{\star}
    \notin
    \mathcal{S}_N
    \}, \\
    K_{\rm sw}
    &\in
    \{K,MK\}, \\
    \rho_{\rm oh}(N)
    &=
    \rho_{\rm fs}
    \frac{N}{K_{\rm sw}}.
\end{align}
\end{subequations}
Define the overhead factor and the miss-penalized raw-rate lower term as
\begin{subequations}
\label{eq:app_eff_factor_group}
\begin{align}
    \zeta_N
    &\triangleq
    1-
    \rho_{\rm fs}
    \frac{N}{K_{\rm sw}}, \\
    \bar R_N
    &\triangleq
    \mathbb{E}[R^{\star}]
    -
    R_{\max}
    \bar p_{\rm miss}(N).
\end{align}
\end{subequations}
Combining \eqref{eq:app_probe_miss_group}--\eqref{eq:app_eff_factor_group}, the effective-rate lower bound and the finite-codebook probing budget are
\begin{subequations}
\label{eq:app_effective_rate_group}
\begin{align}
    \mathbb{E}[R_{\rm eff}^{(N)}]
    &\ge
    \zeta_N\bar R_N, \\
    N^{\star}
    &=
    \underset{1\le N\le K_{\rm sw}}{\arg\max}
    \;
    \zeta_N\bar R_N.
\end{align}
\end{subequations}
Equations \eqref{eq:app_sinr_error_group}--\eqref{eq:app_effective_rate_group} complete the communication chain: sensing uncertainty induces a CRLB-limited SINR-map error, the fused uncertainty yields inverse-uncertainty fusion and robust MCS backoff, and the same uncertainty-margin structure controls the miss probability and effective-rate loss of reduced probing.

\endgroup

\begingroup\scriptsize
\setlength{\itemsep}{0pt}
\setlength{\parskip}{0pt}
\setlength{\parsep}{0pt}

\endgroup

\end{document}